\documentclass{IEEEtran}
\usepackage{cite}
\usepackage{amsmath,amssymb,amsfonts}
\usepackage{algorithmic}
\usepackage{graphicx}
\usepackage{textcomp}

\ifCLASSINFOpdf
\else
\fi

\usepackage{graphics}
\usepackage{graphicx}
\usepackage{epstopdf}
\usepackage{grffile}
\usepackage[dvips]{epsfig}
\usepackage{epsf}
\usepackage{here}
\usepackage{amsthm}
\usepackage{amsfonts}
\usepackage{amsmath}
\usepackage{amssymb}
\usepackage{verbatim}
\usepackage{upgreek}
\usepackage[normalem]{ulem}
\usepackage{soul}
\usepackage{color}
\usepackage[mediumspace,mediumqspace,squaren,binary]{SIunits}
\usepackage{textcomp}
\usepackage{url}

\usepackage{subfigure}
\newcommand{\fref}[1]{Fig.~\ref{#1}}
\newcommand{\sref}[1]{Section~\ref{#1}}
\newcommand{\tref}[1]{Table~\ref{#1}}
\newcommand{\eref}[1]{Eq.~(\ref{#1})}

\def\BibTeX{{\rm B\kern-.05em{\sc i\kern-.025em b}\kern-.08em
    T\kern-.1667em\lower.7ex\hbox{E}\kern-.125emX}}

\begin{document}

\title{On Deep Learning for Radio Resource Management in A Non-stationary Radio Environment}

%\author{\uppercase{Suren Sritharan}\authorrefmark{1},
%\uppercase{Harshana Weligampola}\authorrefmark{1},\uppercase{ and Haris Gacanin}\authorrefmark{2},
%\IEEEmembership{Senior Member, IEEE}}
%\address[1]{University of Peradeniya, Sri Lanka. (e-mail: \{suren.sri, harshana.w\}@eng.pdn.ac.lk)}
%\address[2]{Nokia Bell Labs, Nokia Bell N.V., Copernicuslaan 50, 2018 Antwerp, Belgium (e-mail: harisg@ieee.org)}

\author{\IEEEauthorblockN{Suren Sritharan\IEEEauthorrefmark{1}, 
Harshana Weligampola\IEEEauthorrefmark{1}} and 
Haris Gacanin\IEEEauthorrefmark{2}~\IEEEmembership{Senior Member,~IEEE
\\}
\IEEEauthorblockA{\IEEEauthorrefmark{1}}University of Peradeniya, Sri Lanka\\
\IEEEauthorblockA{\IEEEauthorrefmark{2}RWTH Aachen University, Aachen, Germany}
}

\maketitle

\begin{abstract}
This paper studies practical limitations of learning methods for resource management in non-stationary radio environment. We propose two learning models carefully designed to support rate maximization objective under user mobility. We study the effects of practical systems such as latency and reliability on the rate maximization with deep learning models. For common testing in the non-stationary environment we present a generic dataset generation method to benchmark across different learning models versus traditional optimal resource management solutions. Our results indicate that learning models have practical challenges related to training limiting their applications. The  models need environment-specific design to reach the accuracy of an optimal algorithm. Such approach is practically not realistic since a frequent retraining is needed.
\end{abstract}


\section{Introduction}
\label{sec:intro}
There has been growing interest in deep learning through neural networks (DNNs) for various applications in wireless communications such as resource allocation, channel estimation, interference management, etc. \cite{dl_wireless}. DNN is a supervised learning technique that requires offline data training mechanism with its performance being highly sensitive to the amount and quality of labelled dataset \cite{DL_nature}. Valuable analysis in \cite{dl_wireless} clearly shows that most of the works evaluate the wireless system-level performance under the assumption that environment is stationary, while a training of a learning model is successfully accomplished -- learner is in a steady state with asymptotically converged model. However, these are critical assumptions and their impact needs to be carefully studied.

A misconception of such assumption renders the practical application of deep learning unclear for wireless services in a non-stationary environment with latency and reliability constraints (e.g. Industry 4.0 with moving robots, ultra-dense ultra-range mobile communication networks, mmWave drone-based networks, to name just a few). As we will show in later sections, a non-stationary radio environment causes the ageing of the model with a high dependency on biased and incomplete dataset. This requires frequent retraining of the model and causes a service disruption (e.g. drop of sum rate). Thus, the following questions motivate this study:
($i$) What are the the system's performance limitations with practical training? ($ii$) What is the impact of non-stationarity on learning? ($iii$) What is the efficiency and efficacy of learning with a complexity of the problem?

This paper presents a comprehensive study on deep learning models for resource management in a non-stationary radio environment. We discuss how and by how much a design of the deep learning model is affected by user mobility, problem complexity, retraining time constraints, etc. Unlike previous works \cite{dl_wireless}, we evaluate the wireless system-level performance with respect to the tight latency/reliability learning requirement, and the computational requirements while considering both training and prediction time of learning. Our findings indicate that deep learning models need to be carefully designed to reach the accuracy of an optimal algorithm given the computational complexity constraints, while the latency due to frequent retraining remains a challenge. The contributions of this work are summarized as follows:

\begin{itemize}
\item We present a semi-online training methodology to eliminate service disruptions due to ageing of a learning model. We further enhance our approach with a practical design employing reinforcement learning through Deep Q-Network (DQN) model, to circumvent the frequent retraining requirements through online operation.

\item The sum rate maximization with deep learning in multicarrier systems is prone to power violation. To eliminate the power violation problem, we propose a design of loss function in non-stationary environment for training of DNN models to regulate the power violation.

\item We investigate the learning bias in the subcarrier allocation vector due to its sparsity\footnote{The subcarrier allocation results in an output vector with high number of null values (sparse output).} leads to poor training of the DNN model \cite{DL_book}. To solve this problem we propose a method with linear sequence of learning models referred to as the ``pipeline model''. 

\item Our study reveals that present applications of learning models are not able to support efficient training for real-time applications. They require either novel training approaches or much faster computational resources to reach the latency/reliability criteria of Industry 4.0 applications. Currently available computational resources on handheld devices for learning would only suffice for non-interactive applications.
\end{itemize}

\section{Network Model}\label{sec:problem}

\begin{figure}
    \centering
    \includegraphics[width=0.9\linewidth]
    {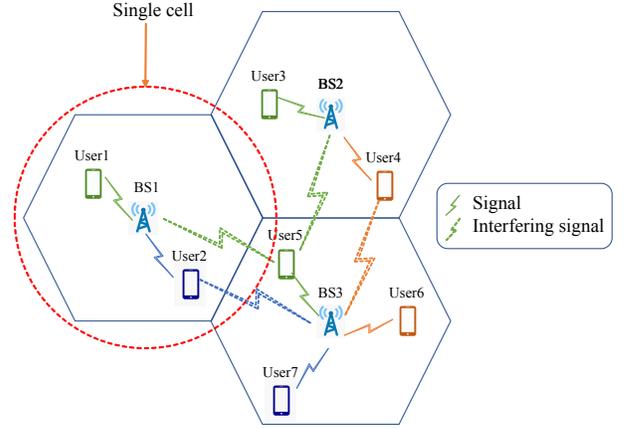}
    \caption{Network model.}
    \label{fig:modelMCMU}
\end{figure}

The study considers a multi-cell multi-user multi-carrier network. \fref{fig:modelMCMU} illustrates a multi-cell multi-user network with $U$ randomly located users. The network consists of $B$ number of base stations (BSs) denoted by the set $\mathcal{B} = \{1,2,3,\dots,B\}$.
We address a particular BS by $b\in\mathcal{B}$ with the set of randomly located users $\mathcal{U}=\{1,2,3,\dots,U\}$.
The users connected to the $b$th BS are given by the set $\mathcal{U}_b\subset\mathcal{U}$, while a user $u\in\mathcal{U}$ is connected to only one BS at a given time, i.e. family of sets $\{\mathcal{U}_b\}_{b\in\mathcal{B}}$ is pairwise disjoint. In OFDM system, a particular subcarrier is denoted by $n$ from a set of subcarriers $\mathcal{N}=\{1,2,3,\dots,N\}$.
%In this system, interference occurs on the downlink transmission between a base station (BS) and a user in two different cells utilizing the same  subcarrier. The sum rate of the system is maximized by minimizing the interference through optimal power allocation (a.k.a the interference management problem).

%\subsection{Channel Model}
As the propagation channel, we assume the time and frequency selective Rayleigh fading channel. The $L$-path discrete-time channel impulse response between the $b$th BS and the $u$th user at the moment $t$ is represented by
\begin{equation}\label{eq:channel}
h_b^u(\tau,t)=\sum_{l=0}^{L_t-1}\bar{h}(l,t) \delta(\tau-\tau_l(t)), 
\end{equation}%http://www.cs.tut.fi/kurssit/ELT-44007/Invocom/p3-7/fading_channel/p3-7_2_1.htm
where $\bar{h}(l,t)$ denotes the $l$th path random complex gain. $\delta(\cdot)$ denotes the delta function and $\tau_l(t)$ denotes the time delay of the $l$th path. $L_t$ indicates the time dependency of the number of paths due to user mobility. $\bar{h}(l,t)$ with $M_t$ scattered waves in the receiver vicinity is defined as zero-mean independent complex variables given by
\begin{equation}\label{eq:pathgains}
\bar{h}(l,t)=\sum_{m=0}^{M_t-1}\exp[\theta(l,m)t+\phi(l,m)],
\end{equation}
where $\theta(l,m)=2\pi(m-r_1)/M_t$ and $\phi(l,m)=2\pi r_2$ ($r_1$ and $r_2$ are random uniformly distributed numbers between 0 and 1). We also assume that that transmitter and receiver terminals are moving with pedestrian speed generating slow fading conditions. We note here that as the users move the channel generator model parameters - $L$ and $M$ - are varying at different time instant $t$. Thus, the non-stationary (dynamic) nature of the environment is determined by user mobility and defined by the number of paths $L_t$ and the number of incoming waves $M_t$ in \eref{eq:channel}. The channel gain $g_b^u(n,t)$ is given as an output of the fast Fourier transform of $h^u_b(\tau,t)$. We consider the signal attenuation to be proportional to the distance using a large scale path-loss in decibel as $\eta = -120.9 - 37.6 \log_{10}{d}/{\bar{d}}$, where $d$ and $\bar{d}$, respectively, denote the distance from the BS to the user and the maximum radius of the BS in kilometers.
% SUREN : \sfrac{d}{\bar{d}

Without loss of generality, we made the following assumptions:

%\begin{description}
\textbf{(A1)} We assume $\tau_0 = 0 < \tau_1<\ldots< \tau_{L_t-1}$ with the $l$th path time delay $\tau_{l} = l\Delta$ where $\Delta=1$ denotes the time delay separation between adjacent paths.

\textbf{(A2)} We assume that expectation term $E[\sum_{l=0}^{L_t-1}|h_b^u(l,t)^2|]$ $= \frac{1-\rho^{-1}}{1-\rho^{-L_t}}\sum_l\rho^{-l}\delta(\tau-\tau_l)$, where $\rho$ denotes the channel decay factor that ensures the total energy of the channel is normalized to unity.

\textbf{(A3)} We assume the Jakes's fading model, where incoming rays constituting each propagation path arrive at a user with uniformly distributed angles \cite{jakes_model}; the normalized auto-correlation function is given by $E[h_b^u(\tau_1, t)h_b^u(\tau_2, t+\varsigma)] = J_0(2\pi f_D \varsigma)$ at delay $\varsigma$ when the maximum Doppler shift is $f_D$\footnote{The auto-correlation function $E[h_b^u(\tau_1, t)h_b^u(\tau_2, t+\varsigma)]$ measures the statistical correlation between two propagation paths with propagation delays $\tau_1$ and $\tau_2$  as a function of time-difference $\varsigma$. $J_0(\alpha) = \frac{1}{\pi} \int_{0}^{\pi} exp(j \alpha cos \theta) d\theta$ is the zero order Bessel function of the first kind.}.

\textbf{(A4)} We assume that the guard interval is sufficiently selected to avoid an inter-symbol interference at any time instant. We assume inter-carrier interference-free system with the block fading, where the fading gains remain constant during the signaling block and vary block-by-block within the frame.

%\end{description}

%represented by 
% \begin{equation}\label{eq:path_loss_dB}
% g_b^u(n,t) = \sum_{\tau=0}^{L(t)-1} h^u_b(\tau,t)exp(-j 2 \pi n \frac{\tau}{N})
% \end{equation}
% for $n\in\mathcal{N}$.

% \Figure[]
% {networkmodel.eps}
% {Network model.\label{fig:modelMCMU}}

Next, we formulate four rate maximization problems that are later used to study the effectiveness of learning models in non-stationary environments.
% SUREN HARSHANA find a editor on journal and put a paper to related work - from march 2019
\subsection{System Model 1: Low complexity case}
\label{sec:CS0_model}
% \hg{Please refer to Fig. 2 in A. B. and C.}
We start with the rate maximization of the single-user single-cell OFDM system (i.e. $U$ = 1, $B$ = 1). The model is described  by a single cell with only one user. The sum rate maximization problem is defined by
\begin{equation}
\label{eq:sr_cs0}
\begin{aligned}
& \underset{p(n)}{\max{}}
& & r = \sum_{n=1}^{N}{log_2[1+\kappa \gamma(n)]} \\
& s.t.
& & 0 \leq \sum_{n=1}^{N} p(n)  \leq  P_{max}
% i = 1, \ldots, m.
\end{aligned}
\end{equation}
where $r$ is the transmission rate of the user. $\kappa = -1.5/\log (5 \, \omega)$ is a constant that determines the transmission rate of M-ary QAM (Quadrature amplitude modulation),
where $\omega$ is the target reliability (i.e. bit-error rate). $\gamma(n)$ is the signal-to-interference plus noise ratio (SINR) between the user and its serving BS on the $n$th subcarrier, $p(n)$ is the power allocated to the $n$th subcarrier and $P_{max}$ is the maximum power allocation of the BS.

The corresponding SINR is given as follows
\begin{equation}
\label{eq:sinr_cs0}
\gamma(n)=\frac{{|g(n)|}^{2} p(n)}{\sigma ^{2}}
\end{equation}
where $g(n)$ is the channel gain between the user and the BS and $\sigma^{2}$ is the variance of the additive white Gaussian noise (AWGN).

To solve the rate maximization problem we utilize the water filling algorithm described in \cite{opt_cs0} as a benchmark for this case study.
%SUREN : check paper again

\subsection{System Model 2: Moderate complexity case}
\subsubsection{Multi-user Single-cell OFDM System}
\label{sec:model_cs1}
The model can be described by a single cell ($B$=1) in \fref{fig:modelMCMU}. The sum rate maximization problem is given by
\begin{equation}
\label{eq:sr_cs1}
\begin{aligned}
& \underset{p(n)}{\max{}}
& & r = \sum_{u=1}^{U}\sum_{n=1}^{N}{log_2[1+\gamma^u(n)]} \\
& s.t.
& & 0 \leq \sum_{n=1}^{N} p^{u}(n)  \leq  P_{max},
\end{aligned}
\end{equation}
where the SINR between the $u$th user and its serving BS, $\gamma^u(n)$, is given by
\begin{equation}
\label{eq:sinr_cs2}
\gamma^{u} (n)=\frac{{|g^{u}(n)|}^{2} p^{u}(n)}{\sigma ^{2}}.
\end{equation}
Here $g^{u}(n)$ is the channel gain between the $u$th user and its serving BS. $p^{u}(n)$ is the power allocation to the $u$th user on the $n$th subcarrier.

The optimization problem consists of two sub-problems namely subcarrier allocation and power allocation. Thus, it is comparatively more complex than the previous case. As the first step of the algorithm, each subcarrier is allocated to a unique user (but a single user can have multiple subcarriers). After subcarrier allocation, power is allocated to each subcarrier. We utilize a greedy method where the subcarrier allocation is done through ranking based on SINR \cite{opt_cs2}. The power allocation is accomplished using the water filling algorithm \cite{opt_cs0} which takes a greedy approach and allocates more power to subcarriers with high SINR. 
% \hg{Guys, any reference for WF? Is this the same reference? }\s{yes, WF is same in both cases (1 and here)}\hg{If it is the same refer to it again. Readers do not know it is the same.}
% Suren : Check paper again

\subsubsection{Multi-user Multi-cell Single-carrier System}
The system is modelled with $B$ BSs and $U$ randomly located users as shown in \fref{fig:modelMCMU}. The sum rate maximization problem is given by
\begin{equation}
\label{eq:sr_cs2}
\begin{aligned}
& \underset{p_b^u(n)}{\max{}}
& & r = \sum_{b=1}^{B}\sum_{u=1}^{U}{log_2[1+\gamma_{b}^{u}(n)]} \\
& s.t.
& & 0 \leq \sum_{u \in \mathcal{U}_b} p_b^u(n)  \leq  P_{max}; \forall b\in\mathcal{B}
\end{aligned}
\end{equation}
where $n$ denotes the frequency component of the SC system. $p_b^u(n)$ is the down-link power allocation between the $b$th BS and the $u$th user. We consider a particular channel realization $h_{b}^{u}(t)$ having $L_t=1$ (i.e. a single path channel) given by \eref{eq:channel}. Now, the SINR is defined by
\begin{equation}
\label{eq:sinr_cs1}
\gamma_{b}^{u}(n)=\frac{|g_{b}^{u}(n)|^{2} p_b^{u}(n)}{\sum_{b'\in\mathcal{B}\char`\\ \{b\}}{|g_{b'}^{u}(n)|^{2} p_{b'}^{u'}(n)} + \bar{\sigma} ^{2}}
\end{equation}
where $p_b^u(n)$ is the downlink power allocation between $b$th BS and $u$th user. $\bar{\sigma}$ denotes the composite Gaussian noise variance of additive noise and residual Gaussian interference after the frequency domain equalization. $g_{b'}^u(n)$ is the channel gain between $u$th user and BSs excluding $b$th BS, $p_{b'}^{u'}(n)$ is the down-link power allocation between BS $b'$ and connected user $u'$. The weighted minimum mean-square error (WMMSE) algorithm \cite{wmmse} is used as a benchmark for this case study. The set difference operator is defined by $``\char`\\"$. %We define except operator $\char`\\$ on sets $A$, $B$ and $C$ such that $A=B\char`\\C$ denotes the elements in $A$ are in $B$ but not in $C$. 

\subsection{System Model 3: High complexity case}
Due to multi-cell interference, the rate maximization problem is Non-deterministic Polynomial-time hard (NP-hard) \cite{convex}. The sum rate maximization problem is given by \eref{eq:sumrate}. 
% The sum rate maximization problem is defined by
\begin{equation}
\label{eq:sumrate}
\begin{aligned}
& \underset{p_{b}^{u}(n)}{\max{}}
& & r = \sum_{b=1}^{B}\sum_{u=1}^{U}\sum_{n=1}^{N}{log_2[1+\gamma_{b}^{u}(n)]} \\
& s.t.
& & 0 \leq \sum_{u \in \mathcal{U}_b} p_{b}^{u}(n)  \leq  P_{max}; \forall b\in\mathcal{B},
% i = 1, \ldots, m.
\end{aligned}
\end{equation}
where $\gamma_{b}^{u}(n)$ and $p_b^u(n)$ denote the SINR and the allocated power between the $u$th user and the $b$th BS on $n$th subcarrier. The SINR $\gamma_{b}^{u}(n)$can be given by
\begin{equation}
\label{eq:sinr_multicell}
\gamma_b^u(n)=\frac{|g_{b}^{u}(n)|^2p_b^u(n)} {\sum_{{b'} \in \mathcal{B}\char`\\ \{b\}}\sum_{u' \in \mathcal{U}_{b'}}\alpha_{b'}^{u'}(n) p_{b'}^{u'}(n) |g_{b'}^u(n)|^2 + \sigma ^{2}}.
\end{equation}
We assume that each user can utilize any of the subcarriers, while users in the same cell cannot utilize the same subcarriers at the same time. The load variable $\alpha_b^u(n) \in [0, 1]$, is defined as the fraction of subcarrier $n$ allocated to user $u\in\mathcal{U}_b$ by time division. Intuitively, $\alpha_b^u(n)$ can be interpreted as the probability of receiving interference from $b$th BS on $n$th subcarrier. To ensure that users in the same cell do not occupy the same subcarrier at the same time, we must have $\sum_{u\in\mathcal{U}_b} \alpha_b^u(n)\leq 1, \forall n \in \mathcal{N}$ and any given period of time.  

Due to the high data rates in BSs, it is not practical to evaluate instantaneous interference and therefore, we consider an average interference taken over time. According to this assumption, the total power %\begin{equation}
%    \label{eq:define_q}
    $q_b(n) = \sum_{u \in \mathcal{U}_b}\alpha_b^u(n) p_b^u(n)$, $\forall b \in \mathcal{B}$, $\forall n \in \mathcal{N}$
%\end{equation}
with \eref{eq:sinr_multicell} leads to the SINR given by
\begin{equation}\label{eq:sinr2}
   \gamma_b^u(n)=\frac{p_b^u(n) |g_b^u(n)|^2 } {\sum_{{b'} \in \mathcal{B}\char`\\ \{b\}}q_{b'}(n) |g_{b'}^u(n)|^2 + \sigma ^{2}}.
\end{equation}
Now, the problem in \eref{eq:sumrate} with \eref{eq:sinr2} is a non-convex optimization problem that we convert to a convex optimization problem and solve by sequential least squares programming (SLSQP) optimization \cite{opt_cs3}.

\section{Related Works}
Deep learning has been studied by many, to address the challenges in wireless physical-layer \cite{dl_wireless}. It plays  an  increasingly  important  role  in the mobile and wireless networking domain. However, to date the deep learning models have been tailored to a specific mobile networking applications as indicated next. 

In \cite{6}, Sun et al. have proposed a DNN model generalization which approximates the WMMSE interference management algorithm with high approximation accuracy and higher computational efficiency compared to state-of-the-art interference management algorithms 
% \s{specify that this results have been shown again in our simulation}.
In \cite{7} Karanov et al. have developed end-to-end deep learning based optical fiber communication which enables the optimization of the transceiver in an end-to-end communication process. The benefits of the proposed deep learning model have been demonstrated by applying it to intensity modulation/ direct detection systems. In \cite{8}, the authors have proposed a resource allocation technique for small cells by employing deep learning for dynamic channel selection, carrier aggregation, and fractional spectrum. In \cite{9} Zhou et al. have proposed an efficient DNN for resource allocation in cognitive radio networks aiming at the real-time performance to maximize the energy and spectral efficiency of the network. However, their work does not highlight the importance and the need for retraining, and thus the performance of the model could degrade in the long run without any retraining. In \cite{HNN} Li et al. have proposed a model that utilizes a Hopfield neural network (HNN) to predict the bit and power allocation in a multi-user OFDM system. Compared to the exhaustive method, the proposed technique is computationally efficient in finding the optimal solution, but the study does not analyze the change in performance with complexity and the ability of the model to perform in a timely manner in the presence of a dynamic environment.

Another approach for solving resource allocation problem is using reinforcement learning methods. DQNs are used commonly in the literature to solve resource allocation for single BS systems. For multiple BS systems, \cite{cs3_dqn} propose a DQN method that uses a DQN model for each BS that is independent of other BSs. 
%We are not going for such an approach because it is a repetition of case study 2. 
We note here that application of DQN method considering information from all BSs requires a large observation to the DQN. In addition, DQN predicts a single action from a predefined discrete set of actions leading to unmanageable action space. Thus, employment of DQN in this multiple BS scenario requires further studies. In what follows, we propose DNN based learning method for resource allocation in this scenario. 
In \cite{cs3_mimo_cnn} authors propose a CNN model to predict power allocation for a Multiple-Input and Multiple-Output systems. However their approach can cater single BS network systems only.

%POINT OUT MISSING PARTS
The studies in \cite{dl_wireless}-\cite{HNN} have neglected the significant practical limitations related to training requirements of the learning model and the impact of ageing of the learning model due to evolution of the wireless environment. Hence, we need to understand the requirements for retraining of the model in the presence of ageing environment and the trade-off between the complexity and the computational efficiency to achieve the target service requirements.

\section{Learning preliminaries}
We first develop a generic framework for dataset generation, and then we briefly describe DNN and DQN models.
%\s{Ageing problem part has been moved to section V}
% \hg{what part?} 
%Finally, we study the ageing problem of the learning model and propose transfer learning to cope with performance loss due to retraining in a non-stationary environment.

\subsection{Dataset Generation Framework}
\label{sec:datagen}
For common testing in the non-stationary environment, we devise the generic dataset generation method that can be used to benchmark across different learning models.

\begin{figure}
\centering
    \subfigure[($L_t, M_t$) pairs]{
    	\includegraphics[width=\linewidth]{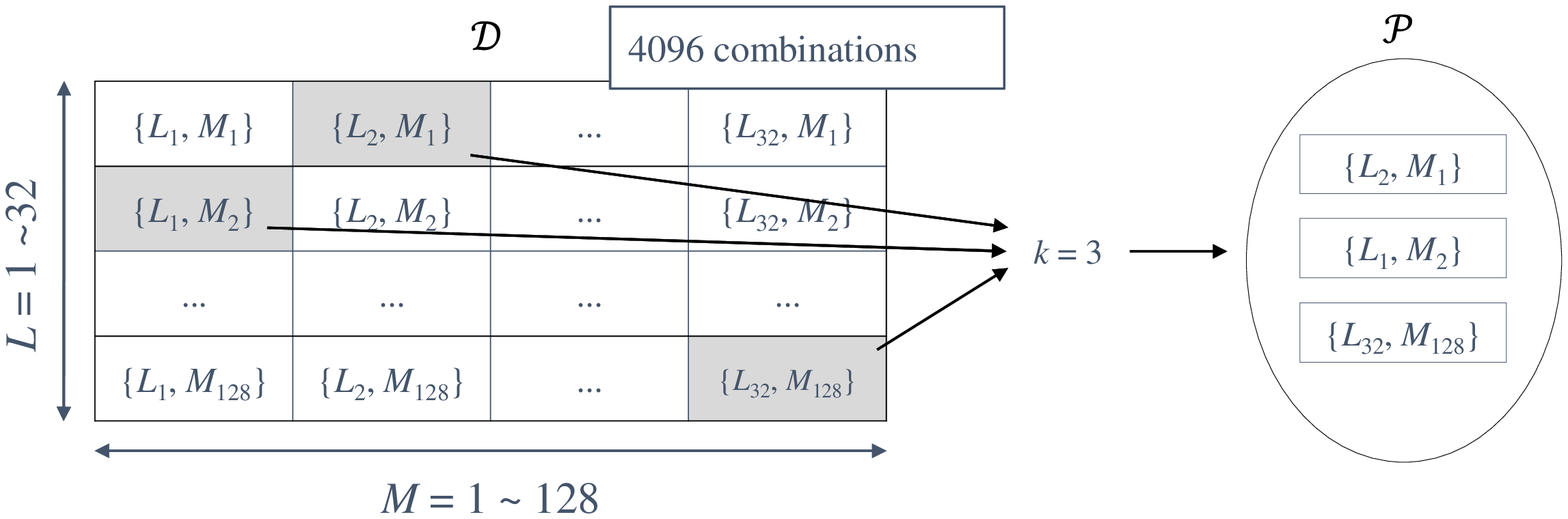}
		\label{fig:datagen_a}
	}
	\subfigure[Framework]{
    	\includegraphics[width=\linewidth]{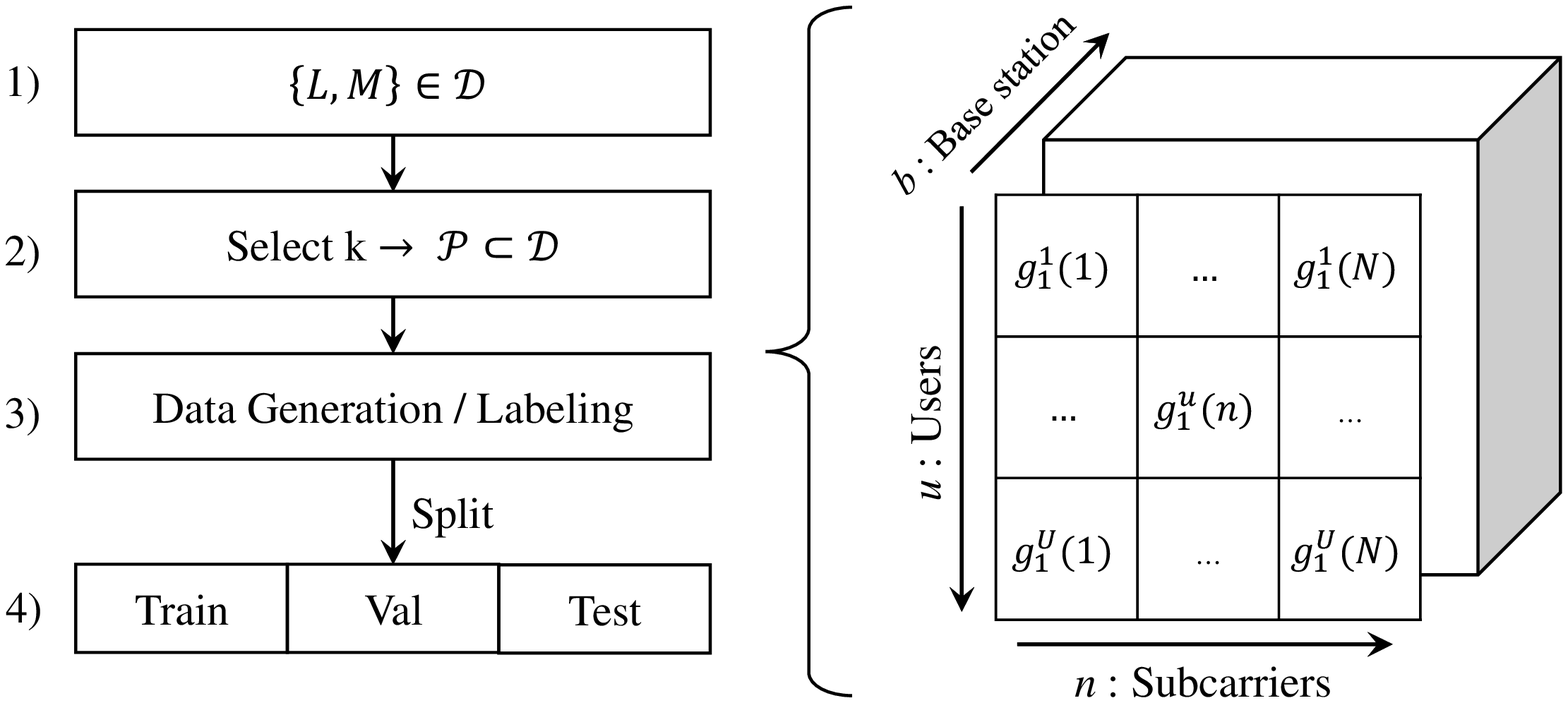}
    	\label{fig:datagen_b}
    }
    \caption{Data generation.}
    \label{fig:datagen}
\end{figure}

\fref{fig:datagen} illustrates the data generation, labeling and training process devised in this paper. This is the general process for data generation method and can be applied to any type of wireless application. The following steps are taken:
\begin{enumerate}
    \item We assume that ($L_t, M_t$) pairs in \eref{eq:channel} vary within a full set 
    %$\mathcal{D}=\{(L, M) \mid L,M \in \mathbb{Z}, 1 \leq L \leq L_{max}, 1 \leq M \leq M_{max}\}$ \s{OR we can use this:}
    $\mathcal{D}=\{(L, M) \mid L \in \mathbb{Z}^+_{L_{max}}, M \in \mathbb{Z}^+_{M_{max}}\}$ where the positive set of integer numbers is defined as $\mathbb{Z}^+_{\alpha} = \{1, \ldots, \alpha\}$ (e.g. if $L_{max}$ = 32, and $M_{max}$ = 128, $L_t$ varies between 1 and 32, while $M_t$ varies between 1 and 128, resulting in a full set with $card(\mathcal{D})$=4096, where $card(\cdot)$ denotes the cardinality of the set. See \fref{fig:datagen_a}).
    \item The non-stationarity due to mobility in the environment is determined by choosing $k$ number of $(L_t,~M_t)$ pairs from the full set $\mathcal{D}$. The non-stationarity factor $k$ defines a subset $\mathcal{P}\subset\mathcal{D}$ ($card(\mathcal{P}) = k$). Non-stationarity refers to the measure of randomness in a dataset. We assume that the users mobility results in an observation of a limited number of of $L_t$ and $M_t$ values which always lie within the subset $\mathcal{P}$; i.e. a higher $k$ describes the higher user mobility and vice versa.
    \item A dataset of $g_b^u(n)$ values are generated as shown in \fref{fig:datagen} by using $k$ selected pairs $(L_t, M_t)\in\mathcal{P}$. Next, the dataset is labelled using a known optimal/sub-optimal algorithm\footnote{Depending on the specific case study the dimensions of the datasample and the specific labeling algorithm may change. The default dataset size is hundred thousand (100K) samples.}. In our sum-rate maximization problem, the labelled dataset describes the output as power allocation for the given input as user mobility determined channel gain.
    \item The labelled dataset is split as Training, Validation and Testing sets\footnote{Unless otherwise stated, the labelled dataset is split as follows: the first 20K data samples are the training set, the next 20K samples are the validation set, and the last 60K are the testing set.} for evaluation of deep learning models as described next.
\end{enumerate}

\subsection{Learning Models}
\subsubsection{Deep Neural Networks}
DNNs are supervised learning models which are used to find structure and patterns in a dataset. Given a labelled dataset with an input and its expected output pair the DNN model learns the dataset structure generators (i.e. pattern) and predicts based on them. For this study, our objective is to predict the optimal power allocation values given a set of channel gain values. At a time $t$ the input is the set of channel gain values 
$\textbf{G}_t = \{g_b^u(n) \mid \forall n\in \mathcal{N}, \forall b \in \mathcal{B} , \forall u \in \mathcal{U}_b \}$ and the target output is the power allocation strategy set given by $\textbf{P}_t = \{p_b^u(n) \mid \forall n\in \mathcal{N}, \forall b \in \mathcal{B} , \forall u \in \mathcal{U}_b \}$. 

\subsubsection*{a) Training}
The training set is used to derive the weights of the model through back propagation. Traditionally, the model is trained such that the loss (cost) function $\Upupsilon$, defined by the mean square error (MSE)\footnote{MSE is generally used as the loss function for a regression problem in a form $\sum_{b\in \mathcal{B}}\sum_{u\in \mathcal{U}_b} \sum_{n\in \mathcal{N}} { \left|\hat{p}_b^u(n) - {p}_b^u(n) \right| }^{2} $ \cite{DL_book}.} between the expected (near-)optimal output $p_b^u$ and the DNN predictions $\hat{p}_b^u$, is minimized. However, the independent allocation of power to each subcarrier might lead to total power being excess of the total power allocation budget $P_{max}$ in \eref{eq:sumrate}, i.e. a power violation problem. This occurs since the DNN model is not aware of the power constraint in \eref{eq:sumrate}. 

\begin{figure}
    \centering
    \includegraphics[width=\linewidth]{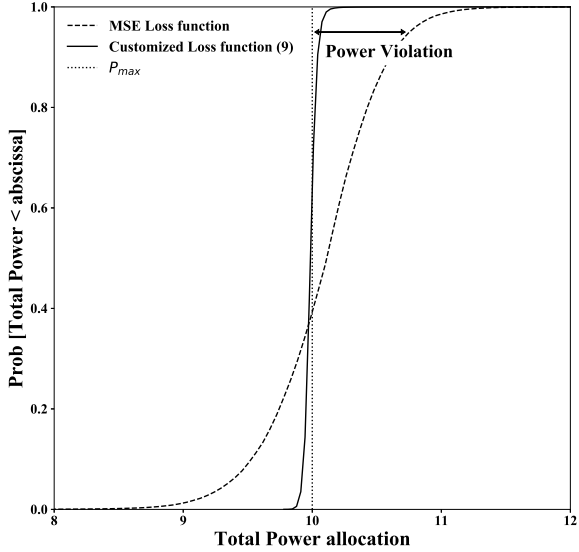}
    \caption{Illustration of power violation problem at the output of DNN model.}
    \label{fig:powerviolation}
\end{figure}

To solve the power violation problem of DNN we introduce a customized loss function\footnote{All DNN models use this loss functions unless otherwise stated.} to regulate the maximum power limit constraint as follows
\begin{equation}
\label{eq:loss_dnn}
\begin{aligned}
\Upupsilon =  \sum_{b\in \mathcal{B}} 
& \left( \sum_{u\in \mathcal{U}_b} \sum_{n\in \mathcal{N}} { \left|\hat{p}_b^u(n) - {p}_b^u(n) \right| }^{2} \right. +\\ 
& \left. \beta \left| \left( \sum_{u\in \mathcal{U}_b} \sum_{n\in \mathcal{N}} \hat{p}_b^u(n) \right) - P_{max} \right| ^{2} \right)
\end{aligned}
\end{equation}
where $\beta$ is the weighing factor of the power violation term.
\fref{fig:powerviolation} shows the effect of the modified loss function and how the standard MSE loss function results in high power allocation which violates the maximum power constraint. Note that the power allocation with traditional MSE loss function \cite{DL_book} exceeds the maximum BS power $P_{max}$ compared to the proposed customized loss function in \eref{eq:loss_dnn}.

The minimization of $\Upupsilon$ is accomplished by varying the model weights, and the minima of the error function is searched by the gradient descent algorithm \cite{DL_book}. The update of weights in each layer independently results in redundant calculation and thus, the back propagation (chain-rule) algorithm is used to avoid redundant calculations. The back propagation is done through the mini-batch gradient descent method\footnote{The training set is broken down into multiple batches in this method. A Batch refers to a set of data samples. Unless otherwise stated, the batch size of 32 samples is used.}.
This results in a higher speed of training due to parallel processing, and avoids problems such as the over/under estimation of the error \cite{batch_GS}.
Furthermore, the Adam optimization technique is used to speed up the training process \cite{adam}. 

The training process is made up of multiple training iterations\footnote{A training iteration is when a single batch has been passed forward and backward through the DNN once}.
After the completion of each training iteration, the updated model is checked for any improvements by the validation stage as described next.

\subsubsection*{b) Validation and Convergence}
After the update of the weights, the improvement in the model are validated in this stage.
% The validation set of unseen data is used to validate the improvements in the updated model.
A model is considered to have improved if the error \eref{eq:loss_dnn} on the validation set has reduced after the update.

Finally the training and validation steps are repeated until convergence for multiple epochs\footnote{An epoch is when the entire training set has been passed forward and backward through a DNN once. In other words, when the training iterations are done on all the batches in the training set, it is known as an epoch.}. The DNN model is said to have converged when the minima of the loss function is reached.
% Convergence of the model depends on the convergence of the loss function.
In general, the back-propagation algorithm cannot be shown to converge, and there are no well-defined criteria for stopping its operation. Furthermore, a low number of epochs will result in under-fitting, while a large number would result in over-fitting \cite{overfit}. Therefore we make a reasonable assumption regarding the convergence point and
% To cope with these issues, we 
terminate the training process by defining the following threshold parameters \cite{earlystop}:
\begin{itemize}
    \item First the change in error $|\Upupsilon_e - \Upupsilon_{e-1}| \, \leq \varpi$ is measure where $\Upupsilon_e$ denotes the error at the $e$th  epoch and $\varpi$ denotes the minimum threshold for the change in error.
    \item The error change is not monotonic, thus an increase in error doesn't indicate that the loss in minimal. Therefore, the change in error measured in ($i$) is monitored for $\hat{E}$ number of epochs and if it is still below $\varpi$, then we assume convergence.
    \item Finally if the aforementioned criteria is not met then the model is stopped after $E_{tot}$ epochs to avoid over-fitting.
\end{itemize}{}

The termination of the training process is therefore determined by either of the three bullets defined above\footnote{Unless otherwise stated, $E_{tot}$ = 100 epochs, $\varpi = 10^{-6}$, and $\hat{E}$ = 50 epochs.}.

\subsubsection*{c) Testing}
Once the DNN model is trained, the sum rate maximization of the wireless system is based on the prediction with DNN model on the test data. Since this data is unseen during the training and validation steps, the evaluation of the model on this set can be generalized to all other datasets.

\subsubsection{Deep Q-learning}
Deep Q-learning is a reinforcement learning model \cite{RL_book, RL_sutton} that combines the deep learning and reinforcement learning to form powerful models which can be used to solve non-stationary problems \cite{DRL_survey}. The DQN interacts with an environment by a set of actions ($\mathcal{A}$ action space) that can be executed, set of environment-specific observable parameters ($\mathcal{S}$ observation space), and a method to calculate the reward after taking a specific action from the action space.
Our objective is to predict the optimal power allocation strategy given a set of observed channel gains. As opposed to DNNs, DQN does not require a labeled dataset. But, DQN can solve only Markov Decision Processes (MDP). If we execute an action by observing the state at $t$ ($\textbf{G}_{t}$) and then observe the resulting state at $t$+1 ($\textbf{G}_{t+1}$), the resulting state is not a consequence of the action taken. Thus, such a process is not a MDP.  Therefore, we focus on time $t$ and reformulate the process of estimating $\textbf{P}_t$ using $\textbf{G}_t$ to a MDP.

\subsubsection{DQN prelimanaries}
\begin{figure}
    \centering
    \includegraphics[width=1\linewidth]{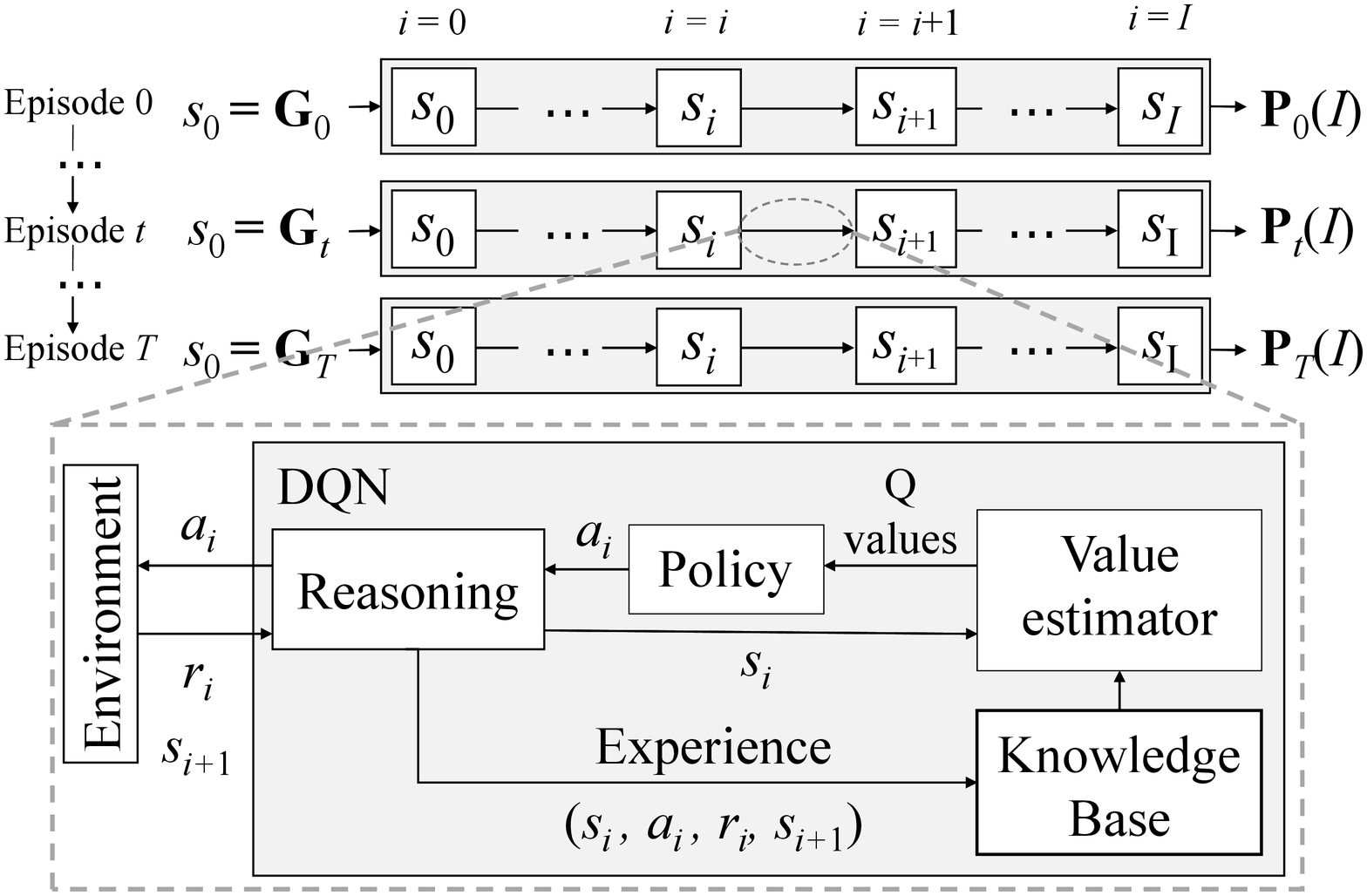}
    \caption{DQN agent model.}
    \label{fig:dqn_model}
\end{figure}

We propose MDP that allows the DQN to adjust the elements of $\textbf{P}_t$. The MDP of DQN estimating the set $\textbf{P}_t$ using $\textbf{G}_t$ for a given time $t$ is defined as an episode in our study. These episodes are repeated at the time $t>[t+1,t+2,\ldots)$ until the system is online. Each episode consist of several steps as illustrated in \fref{fig:dqn_model}. At the beginning of an episode the initial state $s_0$ is observed from the environment. For example, state space can be defined using channel state information (CSI) for each user and subcarrier in a given network. (The exact details about the state space is defined in the relevant section where the DQN model is introduced). Then each state $s_i$ is converted to state $s_{i+1}$ by taking an action $a_i$ on step $i$. This is given by an arrow circled using dotted lines in \fref{fig:dqn_model}. Also, the underlying processes of the DQN are also shown in \fref{fig:dqn_model} and they will be explained later. Estimated power allocation on $i$th step is given by $\textbf{P}_t(i)$.

We propose following MDP processes to estimate power allocation.
 \begin{itemize}
     \item For OFDM systems, in each step $i$ DQN selects the best subcarrier that needs to increase the power by $\delta$ amount from the current power allocation. This power increment of the selected subcarrier is defined as action $a_i$. The DQN selects the action based on the state $s_i$ which is defined by the set $\textbf{G}_t$ with the current power allocation set at step $i$; $\textbf{P}_t(i)$.
     \item For single carrier systems, in the step $i$ DQN selects a discrete power level for a user based on the state $s_i$. The state $s_i$ consist a subset of $\textbf{G}_t$ and $\textbf{P}_t(i)$ that is related to the user.
 \end{itemize}
 
The state $s_i$ is used by the Value Estimator DNN and produce the Q-values for all the possible actions for the state $s_i$. Then, a suitable action $a_i$ is selected using the policy considering the Q-values obtained from the Value Estimator.
Then, $a_i$ is executed on the environment and the corresponding reward $r_i$ for action $a_i$ in state $s_i$ is obtained along with the resulting state $s_{i+1}$ of the environment. (We use the total sum rate of the network given in \sref{sec:problem} as the reward)
Tuple $(s_i,a_i,r_i,s_{i+1})$ is defined as an experience and it is stored in the Knowledge Base.

In a new episode usually DQNs reset the environment to a predefined starting state. But in our case, in a new episode the channel gain $\mathbf{G}_{t+1}$ is correlated to the previous episode's channel gain $\mathbf{G}_{t}$. Thus, there are two approaches to initialize the power allocation at the start of an episode.
\begin{itemize}
    \item We assume that a particular time $t$ is independent of previous time steps $t\in(0,t-1)$. i.e. a episode is independent of previous episodes. Therefore, all elements of $\textbf{P}_{t,0}$ are initialized to zero at the beginning of an episode.
    \item We assume that the power allocated at time $t$ is related to time $t+1$. Thus, we allocate $\textbf{P}_{t}(I)$ to $\textbf{P}_{t+1}(0)$ at the beginning of an episode.
\end{itemize}
An episode ends at $i=I$ when the entire set $\textbf{P}_t(i=I)$ is estimated or if $\textbf{P}_t(i)$ leads to a power violation\footnote{When the sum of power allocated for subcarriers in a certain BS surplus the maximum power of that particular BS, the system is said to be in a power violation state.}. DQN phases are illustrated in \fref{fig:dqn_model} and described as follows

\subsubsection{DQN operation} we describe the operation of DQN through the following four phases:
\subsubsection*{Warmup phase}
DQN agent has no experience about its environment and knowledge base is empty (i.e. observation, taken action and the corresponding reward). Initially, the weights of the Value estimator are initialized randomly in the range [0,1]. To start with training, we obtain experience by executing actions on the environment based on the Q values from the initialized value estimator without training the DQN. Then these experiences are stored in the knowledge base. This phase is known as warmup phase. We select the Warmpup phase to be 1 episode, so that the knowledge base is adequately filled.

\subsubsection*{Exploration/Exploitation Mechanism \cite{RL_sutton}}
The decision-making function (or Policy) is in a dilemma of selecting the action that worked so far - exploiting the knowledge. On the other hand, the policy choose a random action, not considered Q values to gain more reward - exploration for higher rewards. By only exploiting DQN cannot reach the optimal solution, while if we often explore the convergence is slow. The trade-off between the exploration and exploitation is designed by $\epsilon$-greedy strategy in the policy. 
%It choose a random action regardless of the Q values with a probability of $\epsilon$ and the policy choose the action with maximum Q value with a probability of $1-\epsilon$. An action is randomly selected at the beginning (or $\epsilon$ is large) to promote more exploration. This $\epsilon$ is reduced linearly along with the training process to exploit the observed behaviour of the environment. 
DQN starts training phase with $\epsilon$=0.8 at the beginning and linearly reducing to $\epsilon$=0.01 within 1000 episodes and maintained at 0.01 until the system is a online (1000 episodes is equivalent to 1 second since we sample channel gain information for each 1ms). This specific number of episodes is selected after trial and error so that the DQN converges to a solution quicker.

\subsubsection*{Training}
The Q-value for the state $s_i$ and action $a_i$ is given by
\begin{equation}
    \label{eq:q_function}
    Q(s_i,a_i;\theta) = E[r_i + \Gamma Q(s_{i+1},a_{i+1})]
\end{equation}
where $\theta$ denotes the parameters/weights of the value estimator implemented by DNN in \fref{fig:dqn_model}. $\Gamma$ (=0.99) is the discount factor which adds the effect of valuing rewards received earlier higher than the rewards received later.
During training a set of experiences are retrieved from the knowledge base. We train the value estimator to predict Q values such that the DQN maximizes the final reward by updating $\theta$ as follows

\begin{figure}
    \centering
    \includegraphics[width=1\linewidth]{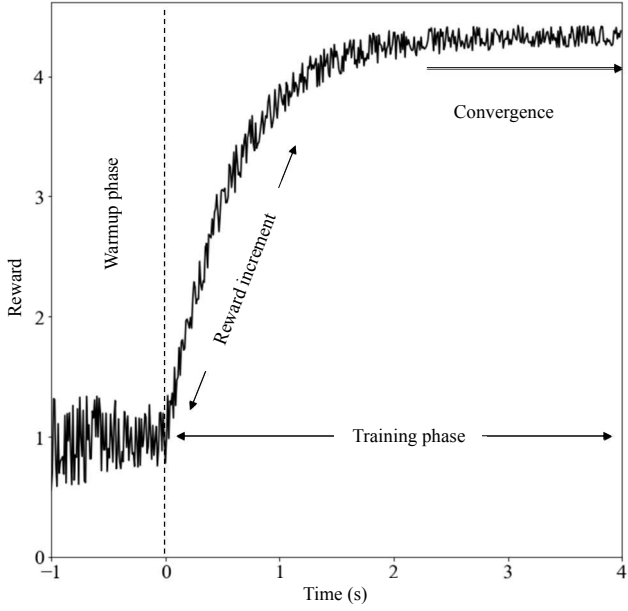}
    \caption{Different phases in DQN training process.}
    \label{fig:dqn_synthetic_rewards}
\end{figure}

\begin{equation}
\begin{aligned}
\label{eq:q_function_update}
\theta_{i+1} := \theta_i + \mu \left[r_i + \Gamma \max_{a'} Q(s_{i+1},a';\theta_i) - Q(s_i,a_i;\theta_i)\right]\\ \times \nabla Q(s_i,a_i;\theta_i)
\end{aligned}
\end{equation}
where $\mu$ is the learning rate (hyper parameter for optimization algorithm that determines the size at each iteration while moving toward a minimum of a loss function).

\subsubsection*{Convergence and Prediction}
When the training phase begins, the DQN starts to train the value estimator to predict Q values that are more likely to give higher rewards. The reward increase in general during the training phase as shown in the \fref{fig:dqn_synthetic_rewards}. This reward increment period can vary with the complexity of the problem, DNN model of the value estimator, hyper parameters used to train the DQN and the computing power. A DQN is said to be converged when the reward given by the DQN for a specific environment plateaued in a maximum reward as illustrated in \fref{fig:dqn_synthetic_rewards}. We can assume that the action taken by the DQN is near optimal after the convergence of the DQN.

Since DQN is a online model, prediction can be done while training. When predicting, $\epsilon$-greedy policy uses $\epsilon = 0$. The reward increases with the experience gained while exploring/exploiting the environment.

\begin{figure}[t]
    \centering
    \includegraphics[width=\linewidth]{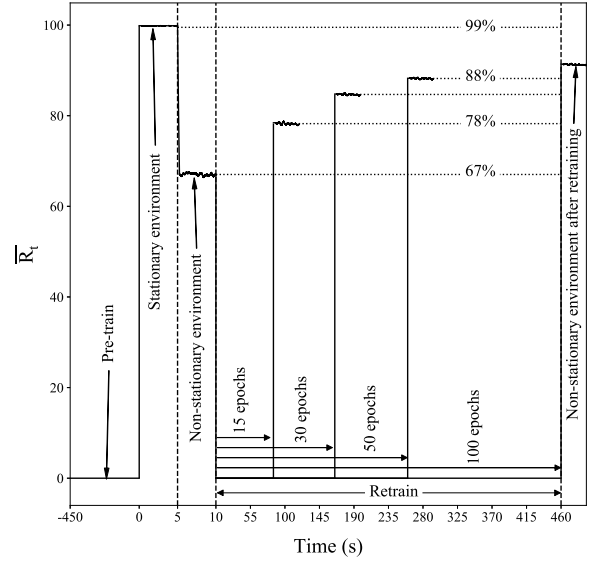}
    \caption{Effective performance of DNN with time averaged for $n = 20$ simulations.
    \label{fig:rel_sr_ma}}
\end{figure}

\begin{figure}[t]
    \centering
    \includegraphics[width=\linewidth]{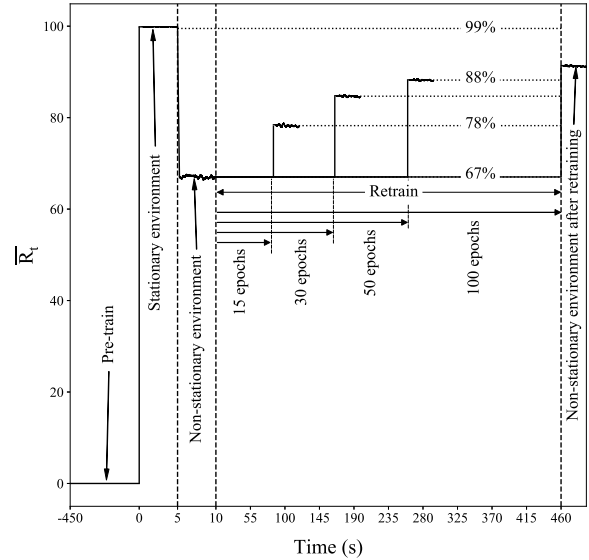}
    \caption{Effective performance of DNN with time with semi-online learning.
    \label{fig:rel_sr_ma_w_transfer_learning}}
\end{figure}

\section{Case Studies and Discussions}
\label{sec:casestudies}

\subsection{Model ageing in a non-stationary environment}
\label{ageing}
Firstly, we focus our attention on ageing of the learning model with respect to non-stationary radio environment. We consider a sum rate maximization problem as described in \sref{sec:CS0_model} for a single-user single-cell OFDM system with $N=32$ subcarriers. 

We consider the environment to be stationary during the initial stage in which the environment is trained. The channel between a stationary user and a BS is defined by \eref{eq:channel} with $L_t=6$, $M_t=20$ scattered waves in the users vicinity, and an exponential power delay profile with decay factor $\rho=0.2$ dB. We utilize  a DNN with 3 hidden layers with a dataset generated as described in \sref{sec:datagen}. The water filling \cite{opt_cs0} algorithm is considered to derive the expected power loading and was used to label and slice the dataset as shown in \fref{fig:datagen}.
% The DNN is trained for 100 epochs with a batch size of 32 samples. 

\subsubsection{Stationary}
The DNN is trained under this stationary environment for 450 seconds using 20K training and 5K validation samples. We consider this time to be the $t=0$ point. Next ,the system performance is evaluated by the relative sum rate defined as
\begin{equation}
\hat{R}_t=\frac{\hat{r}_t}{r_t} \times 100\%,
\label{eq:rel_sr_wf}
\end{equation}
where $\hat{r}_t$ is the sum rate obtained by the output of DNN model and $r_t$ is the expected sum rate given by the water filling algorithm \cite{opt_cs0}. Since the relative sum rate per sample might have a high dynamic range we define the moving average of the relative sum rate as follows:
\begin{equation}
\bar{R}_t = \frac{1}{W} \sum_{i=t-W}^{t} \hat{R}_i.
\label{eq:ma_rsr_wf}
\end{equation}
where $W$ is the time window size\footnote{Unless otherwise stated, a window size of $W=1000$ is employed.}.

\subsubsection{Non-stationary}
Initially, the $\bar{R}_t$ value is $\approx$ 99.5\% indicating that the DNN output for the trained stationary environment is close to the optimal sum rate for $t\in(0,5)$ as shown in \fref{fig:rel_sr_ma}. However, the environment evolves that is partially observed by the trained DNN model and the sum rate drops for $t\in(5,10)$. The model ages in time and retraining becomes an important requirement in a non-stationary environment. Now, the characteristics of the dataset may vary with time. In order to observe the effect of non-stationarity, the user moves at vehicular speed of 20 km/h after the time of 5s (5K data samples) as shown in \fref{fig:rel_sr_ma}.
% \hg{Maybe this would be better to be translated to seconds not K data samples.This could be simply readout from the figure?}\s{It is easier to express the size of this "test" dataset in terms of sample size. This can be converted to time based on the sampling rate of input, but would this make much sense?}
% \hg{Well, the samples of data are not referring to time. 100K samples can be in 10sec or in 10min. I think we should consider time as reference that is shown in nice fig. 3. Otherwise, how can someone relate 100K sample time with time axis in fig. 3.} \s{Discuss: Sampling frequency}

The user movement changes the channel non-stationarity and thus, environment generators such as the distance, multipaths and scattered waves vary as a result. Therefore a new dataset is created with $k=128$, $L_{max}=32$  and $M_{max}=128$ as described in \sref{sec:datagen}.

The relative sum rate of the pre-trained DNN in \fref{fig:rel_sr_ma} was measured under this new environment. In comparison to the optimal solution, a significant drop in the DNN prediction performance is observed due to the non-stationarity after the $5^{th}$ second.
% \hg{between the 5$^{th}$ and 10$^{th}$ second $->$ I do not like how I wrote this. Any suggestions to improve this?}.
This is because, the pre-trained DNN model does not capture the various environment generators defined by $k=128$ due to user mobility. 
% \hg{How about $k=4096$? What is the situation with fully learnied model about its environment? Can we make some comment about that and do we have any graph/table showing this example?} \s{Difference between k=128 and k=4096 is minor. This saturate after a certain point (k ~= 50)}

The DNN model was next retrained after the $10^{th}$ second with data generated in the new environment. The training process was redone mutiple times to analyze the effect of training duration. The training duration (number of epochs) was increased as shown in \fref{fig:rel_sr_ma} and observed that the relative sum rate increases from 78\% to 89\% as the number of epochs is increased from 15 to 50. However, there is no significant improvement when its increased from 50 to 100 epochs. This confirms that the loss function reduces and saturates after sometime which is shown by the increase and saturation of the relative performance. Furthermore, we note that the performance increases with the number of epochs at the expense of the efficiency. This trade-off is studied in detail later.
% Therefore, we conclude that a higher training time results in higher relative performance, but it saturates after a while. \hg{This is because....} 

We note here that the time axis in \fref{fig:rel_sr_ma} has been scaled to clearly visualize the performance and thus the training time is much longer compared to the prediction time.

Another observation is that during the retraining period the relative sum rate of drops to zero. This is because the training set is labelled and the DNN is retrained and thus the DNN model cannot make any new predictions during this time period. This is referred to as the offline training period. In a highly non-stationary system, this offline retraining is a major requirement since it must be done repeatedly as the performance degrades below a threshold. The offline training effectively reduces the overall performance and efficiency of the DNN and is one of the major drawbacks of DNNs.

\subsubsection{Online training through dual DNN in a non-stationary environment}
It can be seen from the \fref{fig:rel_sr_ma} that the relative sum rate $\bar{R}_t$ falls to zero during the training period. This occurs since the described approach and model cannot be used for prediction while training and this degrades the overall effective performance of the model due to frequent training triggered by non-stationary environment. We note here that Training and Validation can be done alternatively. However, the point we are trying to emphasize is that the learning model cannot predict while being trained and consequently, the effective performance goes to zero (i.e. No prediction, thus no output).

To overcome this issue, we exploit the fact that the current model is already trained well enough for the stationary system and therefore. We introduce a semi-online training methodology, whereby the learning model copes with the drop in performance through continuous retraining. However, the model cannot make predictions while being trained. Thus dual DNN models are used in parallel, one for training and another for prediction. The prediction model does continuous predictions, while the training model continuous the cycle of: Reading input, labelling, training and validation and periodic update of the prediction model weights. We maintain the low performance of non-stationary model, while training using the same DNN: we maintain the effective performance of 67\% using this semi-online learning mechanism as shown in \fref{fig:rel_sr_ma_w_transfer_learning}. Consequently, by employing this approach we guarantee a continuous operation of the wireless system without service dropping to zero while maintaining the prediction performance in non-stationary environments. However, the channel changes can occur rapidly and not gradually as studied in this section. This would trigger frequent retraining and such a model requires sophisticated learning system for parallel operation. Therefore, retraining a model in a highly dynamic environment would not be practical due to the limited availability of resources.

\subsection{System Model 1: Low complexity rate maximization}
\label{datagen_cs0}
We consider an OFDM system with $N$ = 128 subcarriers and M-ary QAM modulation. For the propagation channel, we have chosen the normalized Doppler frequency $f_D\times T\times N=5\times10^{-5}$ where $f_D$ = 40 Hz is the Doppler frequency with 1/T = 100 Mbps. Such Doppler frequency corresponds to moving terminal speed of 22 km/h for 2 GHz carrier frequency.\footnote{Unless otherwise stated, we use the same parameters for data generation of all other system models.} The computer simulations were done using $\sigma^2$ = $10^{-3}$ \textmu{\watt}, $P_{max}$ = $10$ \textmu{\watt} and $\omega=10^{-9}$.

A dataset is generated with a non-stationarity factor of $k=10$, then labelled using the water-filling algorithm \cite{opt_cs0} and finally split as shown in \fref{fig:datagen}. We consider the DNN with five hidden layers each with 300 nodes. The input and output layers have $N$ nodes. The nodes for the input layer represent the normalized channel gains $g(n)$, while the power allocation $p(n)$ for the output layer. Each layer uses ReLU (Rectified linear unit) as the activation function except for the output layer which uses Leaky ReLU with a low slope (i.e. gradient = 0.01) to avoid the dying ReLU problem.

\subsubsection{The impact of problem complexity}\label{perf_complexity}
First, the variation of the relative sum rate given by \eref{eq:ma_rsr_wf} is studied as a function of the optimization problem complexity. This is done by altering the number of subcarriers $N$. Consequently, the model needs to learn a more complex features from the dataset. Apart from the sum rate, the variation of the total training time $t_{train}$ and the prediction time per sample $t_{pred}$ with the number of subcarriers $N$ are also given in the \tref{table:cs0_perf_N}. The window size $W$ of \eref{eq:ma_rsr_wf} is equal to the total number of samples in this case.

\begin{table}
\caption{Variation of $\bar{R}_t$ with number of subcarriers = $card(\mathcal{N})$}
\centering
\begin{tabular}{|c|c|c|c|}
\hline
$card(\mathcal{N})$ & $t_{train}$ (min) & $t_{pred}$ (ns/sample) & $\bar{R}_t$ \\
\hline
32 & 5.8 & 27 & 99\%\\
64 & 6.1 & 29 & 95\%\\
128 & 7.7 & 36 & 90\%\\
256 & 9.1 & 38 & 79\%\\
512 & 12.6 & 52 & 39\%\\
\hline
\end{tabular}
\label{table:cs0_perf_N}
\end{table}

We observe that the performance of the DNN decreases as the complexity of the problem increases proportional to $N$. This is because the DNN model needs to be retrained with more richer dataset to cope with the increase of problem complexity. Furthermore, the efficiency, which is measured in terms of the training time $t_{train}$, and prediction time $t_{pred}$ also worsens as the complexity increases. This is attributed to the change in the input/output layer size of the DNN, which results in higher number of weights.

We observe that the performance of the DNN degrades as the complexity of the problem increases. However the performance can be improved to an acceptable level by altering training parameters such as the number of training samples, the number of training epochs or by modifying the DNN to learn more complex data by increasing the number of hidden layers, the number of nodes. However, this can lead to poor efficiency (longer training and prediction time).
The relationship between the performance and the efficiency is discussed in this section.

The performance is measured in terms of the relative sum rate given by \eref{eq:rel_sr_wf} and the efficiency is measured in terms of the training time and the prediction time of the DNN. The time evaluation has three time steps namely, $t_1$ (labeling), $t_2$ (training and validation), and $t_3$ (prediction on the test dataset) as shown in \fref{fig:dynamicity_efficiency}. Each of these simulations were done by keeping the structure of DNN and the dataset fixed while changing one parameter at a time.

\subsubsection*{a) Impact of the Number of Training Samples}
As the first experiment, the training parameter was changed by varying the number of training samples. The variation of the performance and efficiency as a function of the training size is shown in \tref{table:T1Performance}. We observe that the relative sum rate $\bar{R}_t$ increases with training size up to a certain point before it reaches saturation. This confirms the need for a significant amount of training data to for good prediction even in relatively low complexity cases. 

We also observe that the training time $t_2$ is extremely high compared to $t_1$ and $t_3$. This may pose limitations on learning for applications with tight latency and reliability requirements. For example, real-word applications for Industry 4.0 have very low latency requirements for ultra-reliable low-latency communication (URLLC) which make online retraining a challenge \cite{VR}. Specifically, interactive virtual reality applications requires a latency of about 20 ms to avoid distortion and motion sickness and the, currently available online virtual reality applications have a latency of around 100 ms which can only be used for non-interactive applications such as streaming. However, the training time $t_2$ is in the order of minutes which indicates that online retraining cannot be a solution to the ageing problem in highly dynamic real-time applications as discussed before.

We note here, that the experiments were conducted on a 16 Core i7-6700 processor, which runs at 4.00 Ghz and can perform nearly 40 GigaFlops \cite{CPU}. State of the art processors such as the Ascend 910 delivers 256 TeraFlops which is nearly 5000x times faster. Therefore, even though learning models in our studies are constrained by time for real-time applications, powerful AI (Artificial Intelligence) chips coupled with fast communication technologies supported by 5G cloud-based architectures could be used to implement deep learning solutions. The application and robustness of such solutions must be studied through experimentation and application. However, as mentioned before, the processing time must be less than 2 ms for a round trip time of less than 20 ms for online applications such as virtual reality \cite{URLLC}. Therefore, even though the studied learning solutions could be used for non-interactive applications, it remains questionable as to whether frequent retraining can be done within a short period where the radio environment is evolving -- even with the support of computationally powerful devices supported by cloud architecture.

Finally, we consider the prediction efficiency of the learning model compared to the water filling algorithm. The times $t_1$ and $t_3$ are the total times taken by the water filling algorithm and the DNN model respectively to calculate the power allocation strategy. However, since the total elapsed time may not give a good indication of the performance, the time taken per sample by water filling method $t_1'$ and the DNN model $t_3'$ were measured as shown in \tref{table:T2Performance}. It can be observed that the prediction time  $t_1'$ is a few nanoseconds faster compared to the labeling time $t_3'$. However, we note that this observation is specific only to this simpler case study. In later, more complex case studies, we show that learning models can be highly efficient while promising near equal performance when compared to classical algorithms.

\begin{table}
\caption{Variation of performance with number of training samples with $k=10$.}
% \s{k values are different in all 3 cases. In the table k=10. In the figure, k=1 at first and then 2048 at last.}\hg{I do not understand. Which figure you refer to? Also I do not understand what are all 3 cases. Maybe you should explain on our next call. But for sure this needs to be clear from the text.}}
\centering
\begin{tabular}{|c|c|c|c|c|}
\hline
Training set size & $t_1$ (s) & $t_2$ (s) & $t_3$ (s) & $\bar{R}_t$ \\
\hline
1K & 1.25 & 372 & 4.07 & 87.9\% \\
5K & 1.47 & 534 & 3.31 & 88.4\% \\
10K & 1.83 & 744 & 3.15 & 90.1\% \\
20K & 2.62 & 1002 & 2.64 & 90.3\% \\
50K & 4.16 & 2424 & 1.58 & 90.6\% \\
\hline
\end{tabular}
\label{table:T1Performance}
\end{table}

\begin{table}
\caption{Variation of efficiency with number of training samples.}
\centering
\begin{tabular}{|c|c|c|}
\hline
Training set size & $t_1$ (ns/sample) & $t_3$ (ns/sample) \\
\hline
1K & 59 & 51 \\
5K & 58 & 44 \\
10K & 60 & 45 \\
20K & 62 & 44 \\
50K & 59 & 52 \\
\hline
\end{tabular}
\label{table:T2Performance}
\end{table}

\subsubsection*{b) Impact of the Number of Hidden Layers}
The complexity of the DNN was changed by varying the number of hidden layers. The efficiency and accuracy measures are indicated in \tref{table:Layers}. We note here since the time $t_1$ remains unchanged (i.e. the labeling set is not altered), it is omitted from the table. From the table, one can observe that both the training time $t_2$ and the prediction time $t_3$ decrease as the number of layers decreases. However, the accuracy of the model decreases as well. Through these observations, we depict the inability of the DNN model to learn complex functions through the use of simple models. As the number of hidden layers increases both $t_2$ and $t_3$ increase together with the accuracy. However, after a certain point, $t_2$ and $t_3$ keep on increasing while the increment in the accuracy becomes negligible. This shows that the DNN may not be able to perfectly model a problem beyond a certain limit. On the other hand, it should also be noted that by adjusting the complexity, a good balance between the efficiency and accuracy could be achieved.

Therefore, it should be noted that increasing the complexity of the DNN beyond a certain limit could not only reduce its efficiency but would also lead to poor prediction performance.

\begin{table}
\caption{The impact of the number of hidden layers.}
\centering
\begin{tabular}{|c|c|c|c|}
\hline
Number of hidden layers & $t_2$ (s) & $t_3$ (s) & $\bar{R}_t$ \\
\hline
 1 & 564 & 1.51 & 68.2\% \\
 3 & 810 & 2.28 & 88.4\% \\
 5 & 1002 & 2.64 & 90.3\% \\
 7 & 1194 & 3.67 & 91.0\% \\
 9 & 1578 & 4.89 & 90.8\% \\
\hline
\end{tabular}
\label{table:Layers}
\end{table}

\subsubsection{Variation of performance with non-stationarity}
\begin{figure}
    \centering
    \includegraphics[width=\linewidth]{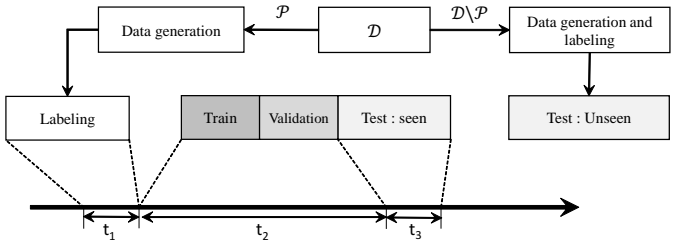}
    \caption{A dataset with seen an unseen $(L_t,~M_t)$ pairs and measuring computational efficiency.}
    \label{fig:dynamicity_efficiency}
\end{figure}

Next the relative sum rate was measured while changing the environment generators of the training dataset; $(L_t,~M_t)$ pairs. Note that in the previous section, the performance of the DNN model was evaluated based on test data which exhibits a low variation when compared to the training data. However in reality, as the channel model varies, the characteristics of the input data would deviate from that of the training data and the performance of the DNN model would change as a result. To observe this variation, a new dataset of 100K samples was generated as shown in \fref{fig:dynamicity_efficiency}. In this case, the $(L_t,~M_t)$ pair values of the new dataset show a high variation compared to the training data. We codify these test sets as seen and unseen data.

\begin{table}
\caption{Variation of $\bar{R}_t$ on seen and unseen data with the channel non-stationarity factor $k$.}
\centering
\begin{tabular}{|c|c|c|}
\hline
$k$ & $\bar{R}_t$ of seen data & $\bar{R}_t$ of unseen data \\
\hline
 1 & 99.5\% & 67.2\% \\
 5 & 93.6\% & 89.8\% \\
 10 & 92.2\% & 89.9\% \\
 50 & 92.1\% & 90.8\% \\
 100 & 91.6\% & 91.5\% \\
%  2048 (50\%) & x\% & x\% \\
\hline
\end{tabular}
\label{table:cs0_dynamicity}
\end{table}

From \tref{table:cs0_dynamicity}, we make the following observations.
\begin{itemize}
    \item Relative sum rate on seen data is high when the known $(L_t,~M_t)$ pairs, $k$ is low. Due to the low non-stationarity factor $k$ the data distribution is narrow and consequently, the DNN model manages to learn the behavior of the data distribution.
    \item As $k$ increases, the relative sum rate on the seen dataset reduces. While increasing the non-stationarity factor $k$ the distribution of the data becomes broader and naturally, the DNN model requires a large set of samples and more parameters to learn the behaviour of the dataset. The number of samples and the DNN architecture is kept constant, thus the overall performance reduces.
    \item The unseen dataset shows an opposite trend, as the relative sum rate increases with $k$. This is because DNN model generalizes the data distribution with more diverse samples. Therefore, we can see an increase in relative sum rate with unseen data with larger $k$.
    \item The relative sum rate of DNN output on both seen and unseen datasets saturates and becomes equal after a certain $k$ value (around 1\% out of 4096). This is because the DNN model reach it's capacity to learn from the given data due to the limitation of the parameters in DNN model.
\end{itemize}

This can be explained through overfitting. As the training data shows low variation, the model easily trains to fit to this training data. This results in a good prediction on the seen data, but the performance on the unseen data suffers as it has a very high variation. However, as the variation of the training data increases, finding a pattern becomes tougher and thus the model generalizes. As a result, the performance on the seen data reduces while due to the generalization, the performance on unseen data increases. This is the main reason for the ageing problem of the model discussed in \sref{ageing}.

% We observe that the performance of the DNN on $Test_1$ is very high when the number of known $(L_t,~M_t)$ pairs, $k$ is low. However, the corresponding $Test_2$ performance is low. This occurs since the DNN model overfits when the variation of training data is low. Now, even though it could predict the known data $Test_1$ with good accuracy, the performance on unknown data $Test_2$ is poor. However, as the dynamicity $k$ value increases, the DNN generalizes the model so that it can predict data with high variance. Now, even though the $Test_1$ accuracy decreases due to the generalization the $Test_2$ accuracy increases. This shows the ability of the DNN model to predict varying data as $k$ is increased. This shows that the DNN generalizes the training as the variation of training data is high. \s{TODO:Rephrase}

\begin{figure*}
    \centering
    \includegraphics[width=\linewidth]{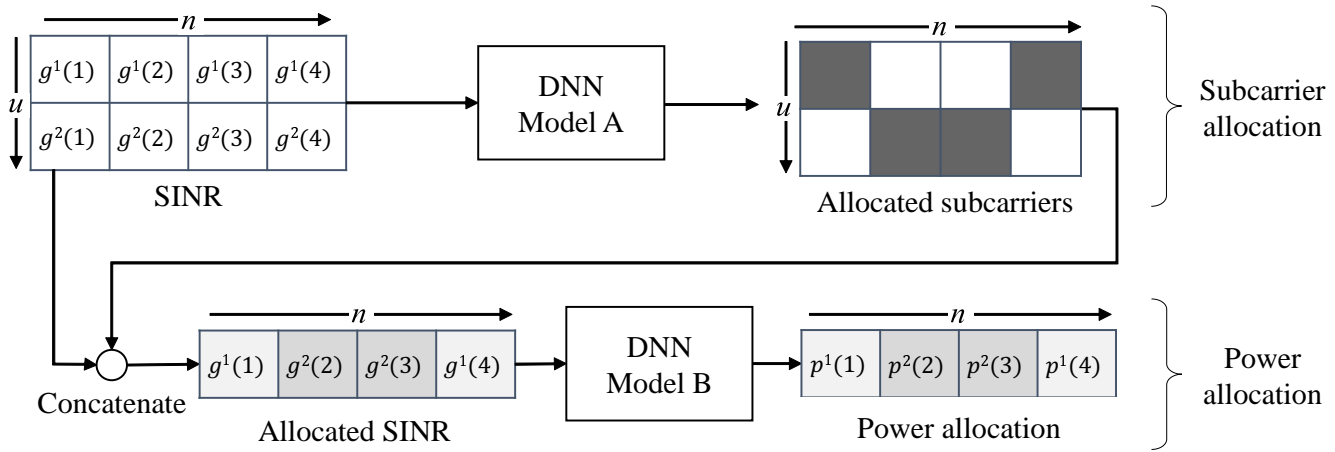}
    \caption{DNN model for single-cell OFDM case.}
    \label{fig:dnn_cs2}
\end{figure*}

\subsection{System Model 2: Moderate Complexity Case}
In this section we consider two scenarios: 1) Multi-user single-cell OFDM and 2) Multi-user multi-cell single-carrier.

\subsubsection{Multi-user single-cell OFDM}\textbf{}
We assume a single cell OFDM system with $N$ = 32 subcarriers and $U$ = 4 users. We assume that there is no interference if the BS does not allocate same subcarrier to multiple users. A dataset is generated with the stationarity factor $k=10$, then labelled using the greedy optimization strategy in \cite{opt_cs2} and split as shown in \fref{fig:datagen}. The data generation process uses the same parameters defined in \sref{datagen_cs0}.

% in this case study is similar to case study 1. The only difference is that the channel gains $g^u(n)$ for each user are generated independently\hg{what do you mean? can we please be more specific here?}. A dataset of 100K samples with $k=2$ is generated and label with the before default split.

As explained in \sref{sec:model_cs1}, each subcarrier is allocated to a unique user and thus, all the other subcarriers have zero power allocation at that time. This leads to a highly biased and sparse set \textbf{$P_t$}. Due to this sparsity, a single DNN model cannot be employed. In such scenario the DNN ``adapts'' to predict 
%\hg{Learns is misleading. DNN does not learn zeros but coefficients. Maybe it would be better to say something like adapts to zeros or something more precise. Also do not use 'x' but ``x''} 
the zeros in the output labels. Thus, converged model predict values close to zero which is not optimal. If we represent sparse data in a $n$-dimensional vector space, they are not clustered closely together. This makes the learning process of the DNN inefficient. To address this problem, we propose a design with two separate sequential DNN models for each sub-problem (subcarrier and power allocation) as illustrated in \fref{fig:dnn_cs2}. DNN model $A$ predicts the subcarrier allocation algorithm, while the DNN model $B$ predicts the power allocation algorithm (water filling).
\begin{itemize}
    \item The DNN Model A has one hidden layer with 200 nodes. The input layer has ($N \times U$) nodes with each node representing the normalized channel gain $g_b^u(n)$. The output layer has ($N \times U$) nodes with each node representing the binary value for the subcarrier allocation. Different approach could be to use lower granularity of CSI such as received signal strength indicator (RSSI) that would impact the performance while the methodology would not change. A true value would mean that a subcarrier is allocated to that user and if the value is false, then it is not allocated. Each layer uses ReLU as the activation function except for the last layer which uses a Sigmoid function. The problem is a multi-class multi-label problem where the users are assumed to be the classes and each user can be labelled with multiple subcarriers. Therefore, binary cross entropy loss function was chosen for the DNN.
    \item Model B predicts the power allocation. Since the labeling algorithm (water filling) for this subproblem is the same as in case study 1, the DNN model is similar as well. The DNN has 3 hidden layers each with 100 nodes. The input layer has $N$ nodes with each node representing the normalized channel gain of the allocated users and the output layer has $N$ nodes with each node representing the power allocation of the particular subcarrier. Each layer uses ReLU as the activation function except for the last layer which uses Leaky ReLU with a low slope (gradient = 0.01) to avoid the dying ReLU problem \cite{dying_relu}.
        %\vcc{Please add a reference about the dying ReLU problem.}
        The DNN uses the same customized the loss function used in the previous study which is given by \eref{eq:loss_dnn}.
\end{itemize}
The combined DNN was evaluated by calculating the average value of the relative sum rate given by \eref{eq:rel_sr_wf} where $\hat{r}_t$ is the sum rate obtained from the DNN output and $r_t$ denotes the optimal sum rate obtained through the greedy optimization approach described in \cite{opt_cs2}.

\begin{figure}
    \centering
    \includegraphics[width=\linewidth]{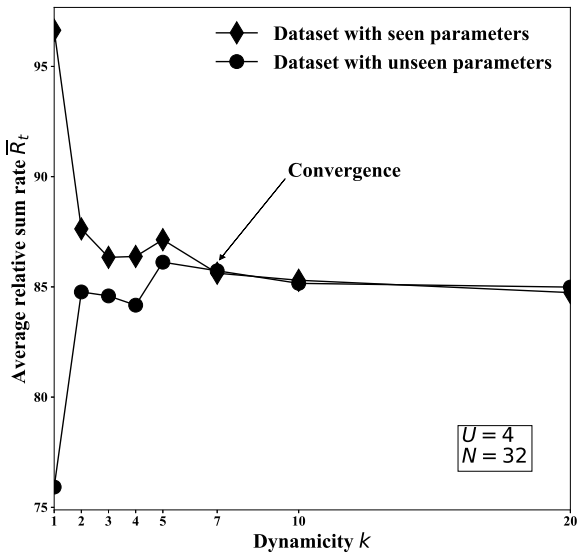}
    \caption{Variation of average relative sum rate with non-stationarity on seen and unseen datasets}
    \label{fig:rsr_k_cs2}
\end{figure}

The variation of the relative sum rate was observed while varying parameters such as the problem complexity, non-stationarity, training parameters (epochs, training set size), and the model architecture. Additionally the change in performance with the number of users is studied and the observations are discussed in the next section.

\subsubsection*{a) Variation of performance with non-stationarity}
% \hg{We have to be specific with title. Title should describe what do we want to emphasize? }

The variation of the average relative sum rate $\bar{R}_t$ with the non-stationarity $k$ is shown in \fref{fig:rsr_k_cs2}. Two examples of datasets with seen and unseen $(L_t,~M_t)$ pair values were generated as shown in \fref{fig:dynamicity_efficiency}. The average relative sum rate $\bar{R}_t$ is low for the unseen data with lower values of non-stationarity factor $k$. This is because the DNN is trained with a dataset with samples from a near non-stationary environment and thus is overfit and biased. For example if we select $k$ = 2, then $\mathcal{D}$ can be $\{(3, 25), (5, 139)\}$. Thus, when the model predicts the power allocation for a data sample which is from a unseen environment with high movement,
% (e.g. $k$ = 2; $\mathcal{D} = \{(23, 456), (31, 346)\}$) 
the performance is poor. However, as the non-stationarity factor $k$ of the training set increases, the DNN model generalizes the behaviour of all dynamic environments and thus, the sum average relative rate $\bar{R}_t$ of seen and unseen datasets converges.

\subsubsection*{b) Impact of the non-deterministic users activation}
When the number of users change, the size of the input vector changes. If the layers in the DNN are fully connected, we cannot change the size of the layer dynamically. Thus, we cannot use the same model. Therefore, we need to train different DNNs that is compatible for different number of users that can be connected to the BS.

If we arrange the channel gain information in a 2D matrix, with users in one dimension and subcarriers in the other, we can assume that the each element in the matrix is spatially co-related. Therefore, we propose to use convolution kernels estimate the power allocation. Thus, we use a Convolution Neural Network (CNN) instead of a DNN to tackle the problem of variation of input users. 
%Harshana: need to add more details

\subsubsection*{c) Recurrent Neural Networks, LSTM cells, to study limitations associated with non-stationarity}
The ageing problem of supervised learning models in non-stationary environment occur due to the dynamic nature of the data. This time varying, dynamic nature can be modelled using memory elements. To study the affect of such networks the DNN was modified by incorporating Long short term memory (LSTM) cells. LSTM cells are a type of recurrent neural networks (RNN) which use "memory cells" to maintain the state information for a longer time period. LSTM cells use complex architecture, which overcomes issues such as vanishing and exploding gradient which are seen in RNNs with simple feedback loops \cite{LSTM}.

The DNN model B (which predicts the power allocation) was modified by adding an LSTM layer at the input. Moreover, RNN models intake past data together with current data. Thus the input data of the neural network was modified such that the input data includes the channel gain values of the previous $10$ time-steps. Our results indicated that after the update of the model with the LSTM cells a similar trend is observed as illustrated by \fref{fig:rel_sr_ma}. This is because the training data still only contains the stationary channel values and the RNN model is biased towards this training data which results in poor performance in non stationary environment. Finally, we conclude that the ageing problem occurs due to the training process. The dependency of supervised learning models on the training data leads to the ageing effect, and this cannot be rectified by modifying the Neural Network architecture.

\subsubsection{Multi-user multi-cell single-carrier} 
We consider a single carrier system with $B$ BSs and $U$ users that are randomly located. The $b$th BS allocates time slots to each user $u \in \mathcal{U}_b$ in a weighed manner in order to maintain fairness. For simplicity, we assume that the $b$th BS always connects to a particular user $u$ and decides the power allocation to minimize interference. Note that due to this assumption, $B$ = $U$ in this case study. However, in the high complexity case we remove this assumption and consider more realistic system. 

A dataset is generated by \eref{eq:channel} with $L_t=1$. In this dataset, a single sample represents $\mathbf{G}_t$; the set of channel gain values $\{ g^1_1, \ldots, g^U_B \}$ between all the users and BSs for a given time $t$. The data are labelled using the WMMSE algorithm \cite{wmmse} and split as shown in \fref{fig:datagen}. We first employ DNNs as the learning model to predict the power allocation.

\begin{figure}
    \centering
    \includegraphics[width=\linewidth]{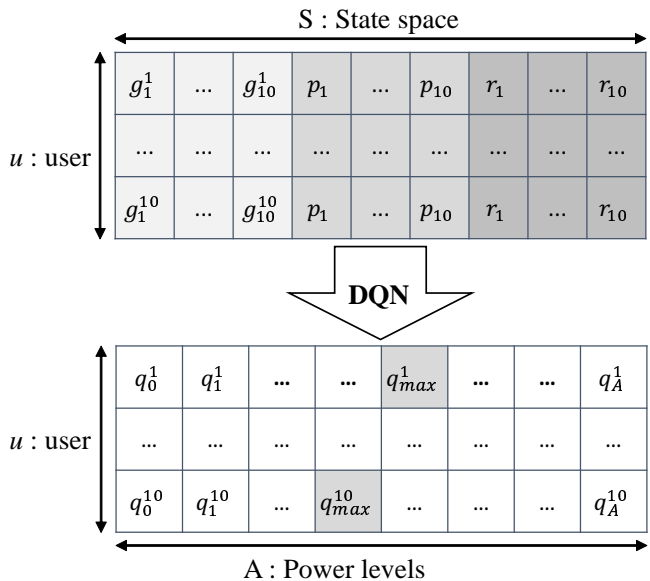}
    \caption{State and Action Space of proposed DQN implementation for Multi-cell Single-carrier system.}
    \label{fig:SA_DRL_cs1}
\end{figure}

\subsubsection*{a) DNN model approach}
To predict the power allocation for each user using channel gain information, we use a fully connected DNN model with three hidden layers having 200, 80 and 50 nodes, respectively. The input layer has $U^2$ nodes with each node representing the normalized channel gain $g_b^u(n)$ and the output layer has $U$ nodes with each node representing the power allocation $p_b^u$ of the particular user. Each layer uses ReLU as the activation function except for the last layer which uses Leaky ReLU with a low slope (gradient = 0.01) to avoid avoid the dying ReLU problem \cite{DL_book}. In a single carrier system the power violation is not an issue since the BS has only one carrier and there is no possibility of allocating power higher than the maximum power allocation. Therefore, we omit the regularization part in \eref{eq:loss_dnn} and thus the loss function is the MSE.

\subsubsection*{b) DQN model approach}
% The results indicated that the performance of the DNN, in terms of both the time consumption and the sum rate, degrades as the complexity of the problem increases. Furthermore, as discussed in previous sections, the ageing problem of the DNN makes it unsuitable under non-stationary channel conditions. Thus we employ DQNs as an alternative approach. Our results indicate that the degradation of DQN's performance with problem complexity is significantly lower compared to DNN's.\hg{Can you please point to concrete results that you mention.} \s{The results (of both DNN and DQN) comes in after this section. Should I move this paragraph down?}\hg{Yes, I think it is better that we move the text where results are shown. Here we should explain what and how we are doing, while in results we should describe what is achieved.}

As discussed in previous sections, the ageing problem of the DNN makes it unsuitable under non-stationary channel conditions. Thus we employ DQNs as an alternative approach. In the proposed DQN design, the power allocation decisions at step $i$ are represented by $\mathbf{P}_t(i) = \{p_1, \ldots, p_B\}$ for all the BSs as illustrated in \fref{fig:dqn_model}.
$\mathbf{R}_t(i)$ represents all the corresponding sum rates $\{r_1, \ldots, r_B\}$ for all the BSs at step $i$.
For this case study we define the state space, action space and reward as follows:

\begin{figure}
    \centering
    \includegraphics[width=0.98\linewidth]{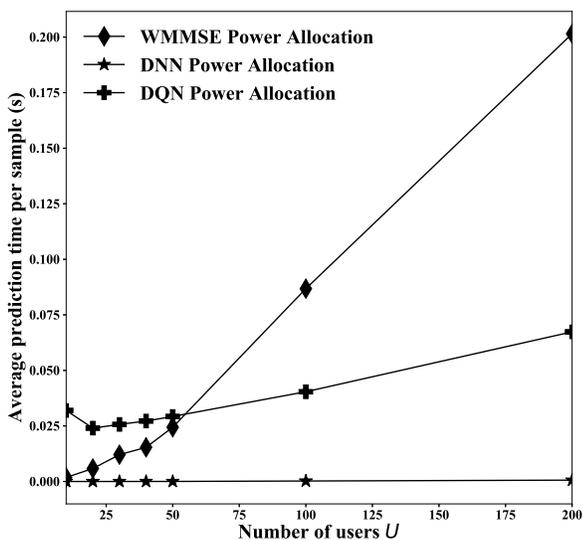}
    \caption{Variation of prediction time with number of users}
    \label{fig:pred_time_cs1}
\end{figure}

\begin{figure}
    \centering
    \includegraphics[width=0.98\linewidth]{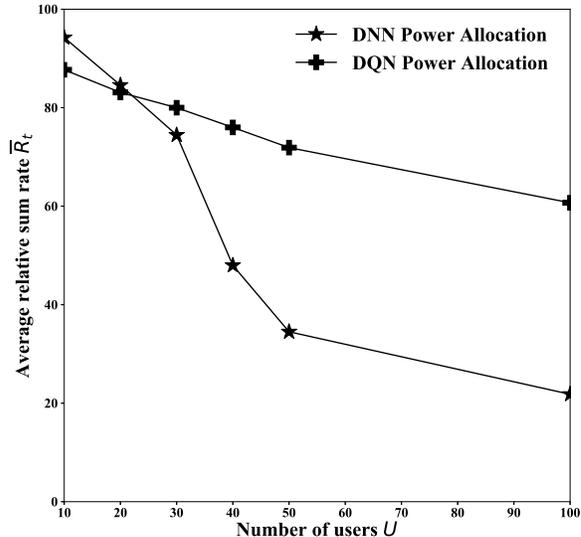}
    \caption{Variation of relative sum rate with the number of users}    \label{fig:rel_perf_cs1}
\end{figure}

\begin{itemize}
\item \textit{State Space} - The state space for a $U$ = 10 system has the values concatenated/stacked together is shown in \fref{fig:SA_DRL_cs1}. The space is composed of the current channel gain values of each user and interfering users $\mathbf{G}_t$, previous power allocation for each user $\mathbf{P}_t(i-1)$, and previous rate of each user $\mathbf{R}_t(i-1)$.  We can use this state space to predict an action that will estimate power levels of all BSs. But this will require an action space of size $A^U$. This is not practical for DQNs since the action space increases exponentially. Therefore, in a single step $i$ we observe the state of a single user (i.e. a row in state space as shown in \fref{fig:SA_DRL_cs1}). Let's call this the observable state. Using this observable state we predict the BS power level for that particular user using the DQN. Similarly we predict the power level for all the users in the step $i$. After that we proceed to step $i+1$. By this way we can reduce the size of state space to $U \times 3$ and action space to $A$.
\item \textit{Action Space} - The action space of a DQN is finite. However, since the power allocations in our case can have infinite values, the total power was divided into $A$ discrete power levels. \fref{fig:SA_DRL_cs1} shows the action space for $U$ = 10 users.

\item \textit{Reward} - We use the sum rate which is given by \eref{eq:sr_cs2} as the reward for the DQN.
\end{itemize}

Finally, the DNN used as the value estimator of the DQN as illustrated in \fref{fig:dqn_model} is a fully connected neural network. It has 3 hidden layers each with 128 nodes. The input layer has ($U$ x 3) nodes. The output layer has $A$ nodes. Each layer uses ReLU as the activation function except for the last layer which uses the linear function. The DQN uses the MSE as the loss. The policy used for this model is the epsilon greedy policy with vanishing epsilon value from 0.8 to 0.001.

Our discussions are centered around the two metrics: the relative sum rate and the prediction efficiency of learning models. The relative sum rate is given by \eref{eq:rel_sr_wf}, where $\hat{r}_t$ denotes the sum rate $\textbf{R}_t(I)$ calculated using the DQN output $\mathbf{P}_t(I)$ which is illustrated in \fref{fig:dqn_model}, while $r_t$ denotes the expected sum rate given by the WMMSE algorithm. The efficiency is defined by the time taken to predict the power allocation of a unit sample. The variation of the relative performance and the prediction time with the number of users is shown in \fref{fig:rel_perf_cs1} and \fref{fig:pred_time_cs1} respectively.

\subsubsection*{c) Efficiency of learning models}
As shown in \fref{fig:pred_time_cs1}, the prediction times of methods increase as the complexity of the problem increases. The increment in DNN is negligible. The increment in prediction time of the DQN model is slightly higher, whereas the increment of WMMSE time is exponential.
This exponential increase in the prediction time of WMMSE algorithms can be attributed to its iterative optimization technique to find the solution.

On the other hand, DNN takes comparatively less time to predict power allocation after training the model, which depends only on number of parameters in the DNN model.
Therefore, from the figure, we conclude that WMMSE is very inefficient compared to other models.
However, since DNNs require a labeled dataset to be trained, the training time of DNNs would be increased due to the inefficiency of the WMMSE algorithm (since the dataset would have to be labelled first by WMMSE). 

Therefore, DQNs act as a suitable alternatives since they do not require to be pre-trained as DNNs and the increment in the prediction time with problem complexity is low compared to WMMSE algorithm. Next the relative performance of these models with respect to the WMMSE algorithm is analyzed.

% However, due to its poor performance at high complexity it is not a suitable alternative.

% Since the WMMSE algorithm is inefficient, DNNs would take a longer time to train .

% While creating large datasets is impractical, the DNN is trained with smaller datasets and the performance consequently degrades. To address this problem we introduce DQN model.

\subsubsection*{d) Relative sum rate of learning models}
From \fref{fig:rel_perf_cs1}, we observe that, as the number of users increases, the performance of both these models degrade. However, the degradation of the performance of DNN is significantly higher. This occurs because, when number of users increase, the complexity of the problem also increases. However, the DNN model has the same complexity and thus may not be able to model more complex problem. This shows the inability of a DNN to learn a model for a problem correctly. Thus a complex model would be required as the problem complexity increases. However, as discussed in previous section this would make the DNN inefficient and would not be a suitable solution.

As opposed to DNNs, DQNs show low degradation of performance as the problem complexity increases. To identify the reason behind this, the variation of the sum rates throughout the DQN training process is studied in the following section.

\begin{figure}[t!]
\centering
    \subfigure[First 1K episodes.]{
    	\includegraphics[width=\linewidth]{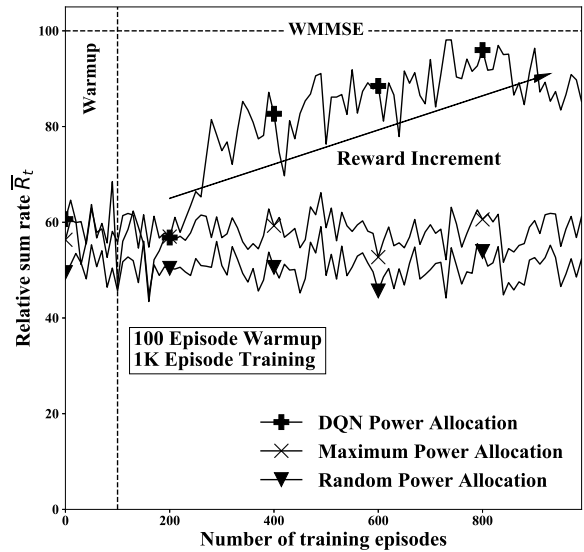}
		\label{fig:DRL_1k_cs1}
	}
	\subfigure[100K episodes.]{
    	\includegraphics[width=\linewidth]{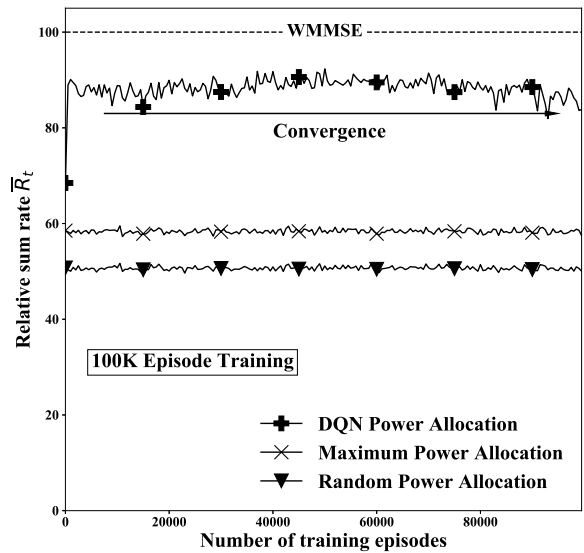}
    	\label{fig:DRL_100k_cs1}
    }
    \caption{Variation of Relative sum rate of DQN throughout the training.}
    \label{fig:drl_perf}
\end{figure}

\begin{figure}
    \centering
    \includegraphics[width=0.97\linewidth]{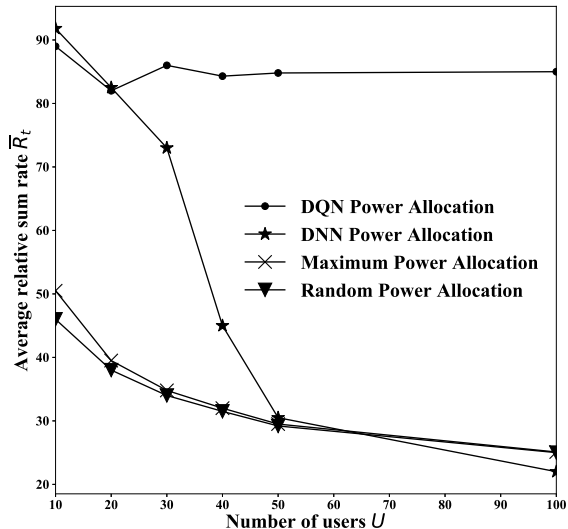}
    \caption{Variation of relative sum rate with number of users after convergence.}
    \label{fig:fin_perf_cs1}
\end{figure}

\begin{figure*}
    \centering
    \includegraphics[width=.8\linewidth]{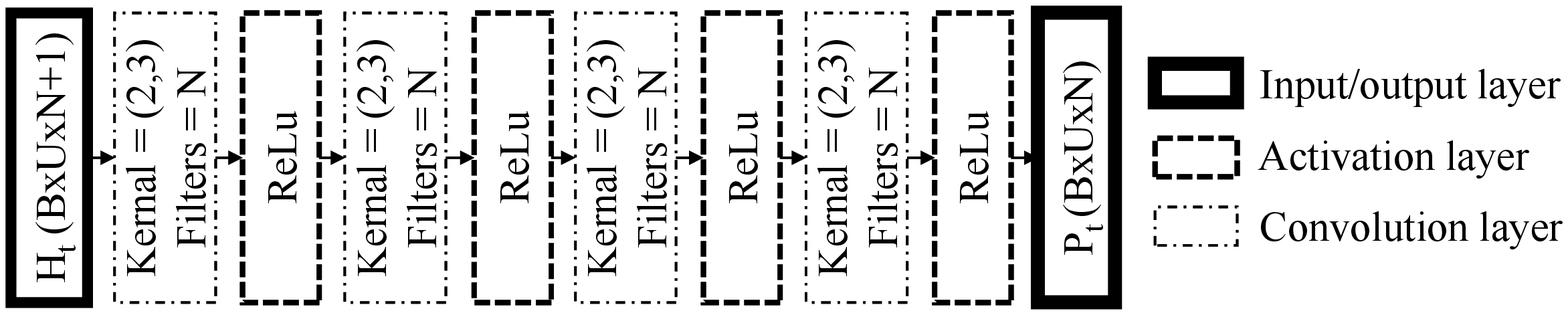}
    \caption{Proposed DNN model to predict power allocation of multi-cell OFDM case.}
    \label{fig:dnn_cs3}
\end{figure*}

% We observe that the relative sum rate degradation of the DNN is much higher than the DQN when the number of users increases. This is because the DNN has a fewer data (experience limited) when compared to continuous training process of DQN through online data. When considering the efficiency of the learning model, if the number of users is low, DQN requires more time to train and predict power allocation than WMMSE. This is because when the number of users is low, WMMSE can quickly find an optimal power allocation, while DQN requires several episodes to train the model and predict an near optimal power allocation. On the other hand, as the number of users increases, DQN becomes more efficient than WMMSE since it learns the behavior of the environment and online trains through time with new data. This allows the DQN to predict near optimal power allocation with lesser time than the WMMSE or DNN. SUREN : TODO : Include these points

% We observe that performance of DNN decreases as the complexity of the problem increases. This is because the DNN requires large datasets to train more complex problems. 

% \hg{Are there any observations and conclusions related to DQN? This and the following subsections are highly disproportionate.}

% DNN degrades. while the prediction time increases. This is because the size of the input vector is increasing. 

\subsubsection*{e) Relative sum rate of DQN during training}
The performance of the DQN is compared with the optimal performance given by the WMMSE algorithm. Furthermore, the reward (sum rate) from random power allocation and maximum power allocation strategies are also considered for comparison\footnote{In random power allocation strategy, power is allocated to each user randomly. In the maximum power allocation strategy, the maximum power of each base station is assigned}. The variation of the reward throughout the training process is shown in \fref{fig:DRL_1k_cs1} and \fref{fig:DRL_100k_cs1}. ``Warmup'' phase is defined by the first 100 episodes followed by the training phase. Note that at the warmup phase, the relative performance of the DQN is nearly equal to the random power allocation method. However, after the saturation (convergence) phase the relative performance of the DQN is much higher than the random / maximum power allocation ratios. This confirms that the DQN has been adequately trained.

We note here that the DQN is initially untrained and would perform poorly compared to the DNN and WMMSE algorithms. Therefore, when the average performance of the DQN is compared with other models as in \fref{fig:rel_perf_cs1}, this would also include the warmup phase and the reward increment phase which have lower performance compared to the saturation value. So, in order to compare the optimal performance of these models, in the next experiment, the DQN was initially trained till convergence. Then the performance of these models was measured on a new dataset.

\fref{fig:fin_perf_cs1} shows the relative performance of the converged DQN together with DNN, random power and maximum power allocation techniques. From \fref{fig:fin_perf_cs1}, we can observe that the final performance of the DQN stays constant as the number of users increases. 
% HARSHANA
This signifies that the DQN would require a higher number of episodes to train as the complexity of the problem increases resulting in a lower average performance. 
Furthermore, we also  observe that after a certain point the DNN performance degrades to a point below the random power allocation performance which shows that DNN would not be suitable for high complexity problems.

These results indicate that the performance of the DNN, in terms of both the time consumption and the relative sum rate, degrades as the complexity of the problem increases. On the other hand, eventhough the WMMSE algorithm provides optimal performance, it has poor efficiency at higher complexity cases. Thus, we can conclude that DQNs gives a good tradeoff between performance and efficiency when compared to the iterative optimal algorithms. 

% degradation of DQN's performance with problem complexity is significantly lower compared to DNN's.

\subsection{Case Study 3: High complexity case}
% - Multiuser Multicell OFDM
We assume a multi-user multi-cell OFDM system with a varying number of subcarriers $N$ having $B$ BSs and $U$ radonmly located users. In this scenario we first consider a non-stationary system and then, we explore the effect of stationarity due to user mobility as described in \sref{sec:problem}. The data generation from the single-cell OFDM case is extended to multi-cell scenario. That is, we calculate the channel gain for each subcarrier between a given BS and all users in the wireless network (including users in other BSs). Then we have a 2D matrix of channel gain information $\mathbf{G}_t$ of size $U \times N$ and calculate the channel gain for all the BSs, stack them to form a 3D matrix $\mathbf{G}_t^3$ of size $B \times U \times N$. Each user is assumed to be connected to the closest BS. The users connectivity is represented in a adjacency matrix $\mathbf{A}_t$ of size $B \times U$, where element $\mathbf{A}_{t,b,u}$ = 1 if user $u$ is connected to the BS $b$ and 0 otherwise. The dataset consist of 1000 samples and each sample consist of channel gain information ($\mathbf{G}_t^3$) and BS-user connectivity information ($\mathbf{A}_t$) for a given time $t$.

\begin{figure}
    \centering
    \includegraphics[width=1\linewidth]{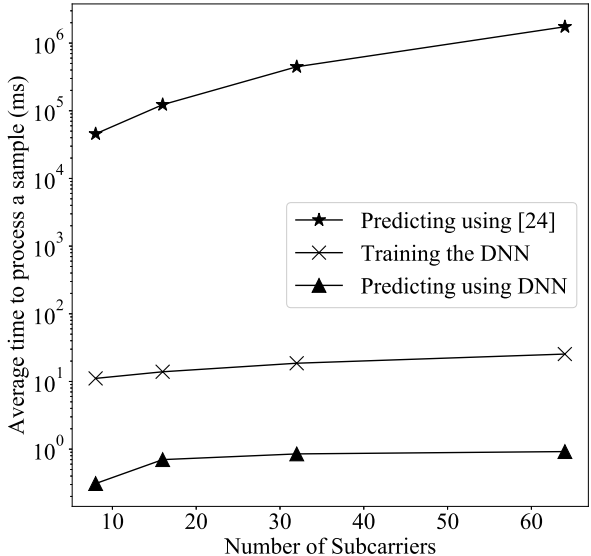}
    \caption{Variation of labeling, training and prediction time with number of subcarriers.}
    \label{fig:dnn_cs3_time}
\end{figure}

\begin{figure}
    \centering
    \includegraphics[width=1\linewidth]{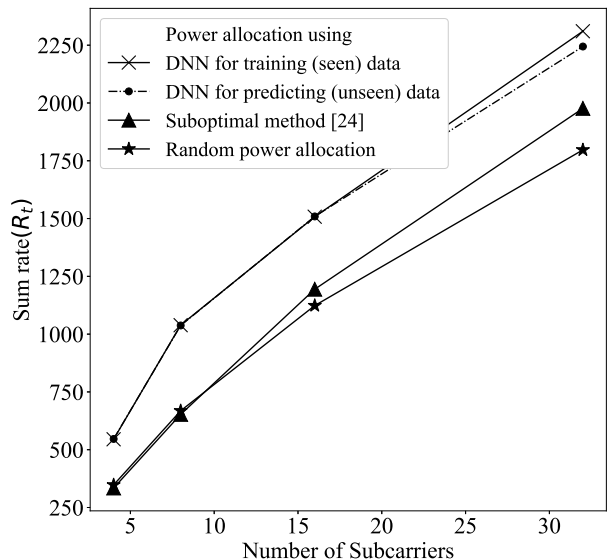}
    \caption{Variation of average sum rate with number of subcarriers.}
    \label{fig:dnn_cs3_rel}
\end{figure}

For each user in the system we need to predict two variables: ($i$) power allocation, and ($ii$) subcarrier allocation. In this case study we define a single DNN model to predict power allocation $\mathbf{P}_t$ and assume subcarriers are allocated equally to all users. Current mobile systems use a distributed system where each cell's decisions are taken independently of their neighbors CSI. However, in 5G NR (New radio) Central Unit and Distributed Unit functional split architecture allows for coordination for performance features, load management, real-time performance optimization. 

We propose a DNN model that inputs $\mathbf{G}_t^3$ and $\mathbf{A}_t$ from input layer as illustrated in \fref{fig:dnn_cs3}. We concatenate $\mathbf{A}_t$ along the dimension of the subcarriers in $\mathbf{G}_t^3$ to construct the 3D input matrix. Thus, the size of the input matrix is $B \times U \times (N+1)$. The proposed DNN model consist of four convolution layers with kernel size $(2,3)$ and $N$ filters. Each layer is passed through a ReLu activation function to make the input-output correlation non-linear. The model predicts power allocation $\mathbf{P}_t$ from the output layer as shown in \fref{fig:dnn_cs3}. We propose a loss function that does not require labeled data to train. Thus, the proposed DNN is non-supervised. The loss function consist of the sum-rate and the power regulation constraint introduced in \eqref{eq:loss_dnn}. The loss function is given as follows,

\begin{equation}
\label{eq:loss_dnn_cs3}
\begin{aligned}
\Upupsilon = \sum_{b\in \mathcal{B}} \left( -\sum_{u\in \mathcal{U}_b}\sum_{n\in \mathcal{N}}{log_2[1+\gamma_{b}^{u}(n)]} \right.+ \\
\left. \beta \left| \left( \sum_{u\in \mathcal{U}_b} \sum_{n\in \mathcal{N}} \hat{p}_b^u(n) \right) - P_{max} \right| ^{2} \right)
\end{aligned}
\end{equation}

Where $\gamma_{b}^{u}(n)$ is given by \eqref{eq:sinr2}. Next, we discuss the average time taken to compute the power allocation and then, we present how the sum rate varies with number of subcarriers. Then, we compare our method with suboptimal solution \cite{opt_cs3}, random power allocation strategy and water filling power allocation strategy.

\subsubsection{Variation of efficiency of the proposed DNN}
The near optimal solution given in \cite{opt_cs3} was used to benchmark the dataset for wireless network where $B$ = 2, $U$ = 4 and $N$ = 16. The reason for selection of lower number of users and BSs is the inability to calculate power allocation for a large datasets using existing near-optimal solutions in a reasonable time. In \fref{fig:dnn_cs3_time} consider the case where there exist $N=16$ subcarriers. DNN takes around 0.7 ms to predict the power allocation per sample in the dataset, while it takes around two minutes per sample with \cite{opt_cs3}. The dataset consist of 1000 samples. Thus for the benchmarking process itself takes around 2$\times$1000$	\approx$33.3 hours. This is because \cite{opt_cs3} takes multiple iterations to converge to a solution which is computationally exhaustive. To train the DNN, it takes around 14 ms $\times$ 1000 = 14s per epoch.
%Exponential time in labeling when compared with predicting time of DNN is because the DNN forward propagate the input vector in the DNN model and get the output from a single pass through the model.

When the number of subcarriers increases, the time taken for \cite{opt_cs3} algorithm increases exponentially as illustrated in \fref{fig:dnn_cs3_time}. This is because when the number of subcarriers increases, the \cite{opt_cs3} algorithm takes more time to compute each iteration because of the large vector representing the state. This implies that when the number of BSs and users increases, the time taken for \cite{opt_cs3} increases exponentially as well. For example, in this wireless system it takes around 1 minute to predict power allocation using a sample from the dataset for a given time $t$. For the system with $N=64$ subcarriers having $B=2$ BSs and $U=16$ randomly located users, it takes around 1 hour to predict a sample. Thus, in order to predict the complete dataset it will take 1000 hours or roughly 41 days. 
As a result, using the method in \cite{opt_cs3} for large wireless systems with multiple BSs and/or crowded areas is not practical because of the time for computation.

\subsubsection{Variation of average sum rate with number of subcarriers}
\begin{figure}[t!]
\centering
    \subfigure[Maximum sum rate]{
    \label{fig:cumsum_capacity_cs3}
    \includegraphics[width=1\linewidth]{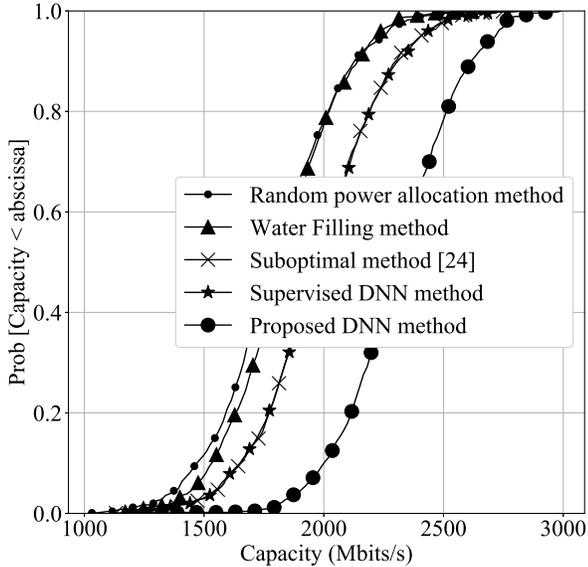}
	} 
	\subfigure[Total power allocation]{	
    \label{fig:cumsum_power_cs3}
	\includegraphics[width=1\linewidth]{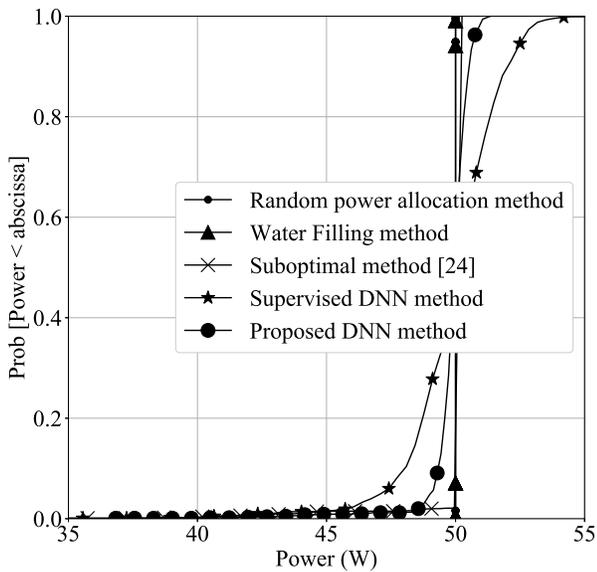}
	}
	\caption{Cumulative probability distribution of sum rate and total power allocation for each BS using random, water-filling, sub-optimal and proposed learning method with DNN (Maximum power allocation for a BS is 50W).}
    
\label{fig:cdf}
\end{figure}

We compute average sum rate using random power allocation, suboptimal method and proposed DNN. \fref{fig:dnn_cs3_rel} illustrates the average sum rate of multi-user multi-cell OFDM with $B=2$ and $U=4$, while the number of subcarriers $N$ varies from 4 to 32. We compare the average sum rate using the training data and prediction data for the proposed DNN. Where prediction data is not used for training. When the number of subcarriers increases the average sumrate increases in all methods since we keep the bandwidth of a subcarrier constant. Therefore, larger the number of subcarriers, higher bandwidth to transmit data. But it is clear that the proposed method gives higher sumrate regardless of the number of subcarriers. When using unseen data, the proposed method decreases in performance when number of subcarriers is equal to 32. This is because, 

We have compared the cumulative sum of the probability distribution of sumrate and total power from different algorithms. The system is defined as $B$ = 2, $U$ = 4, $N$ = 32. \fref{fig:cdf} illustrates the sum rate and total power variation using the proposed DNN, water filling algorithm, random power allocation and suboptimal solution in \cite{opt_cs3}. 
We used water filling algorithm from case study 2 and apply it for each BS independently. Our proposed method improves the sum rate in comparison with other methods. This is because our proposed method uses a unsupervised learning model that consider the all CSI from all BSs. Also, it can be seen that the the proposed method violates total power allocation constraint less frequently as given in \fref{fig:cdf}\subref{fig:cumsum_power_cs3}. This is because the proposed cost function penalizes when power violations occur.

\begin{figure}[t!]
    \centering
	\includegraphics[width=1\linewidth]{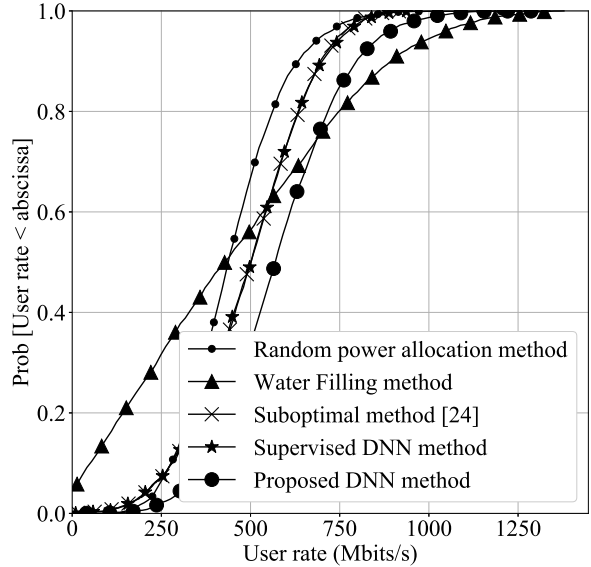}
    \caption{Cumulative probability distribution of sum rate of a user for each BS using random, water-filling, sub-optimal and proposed learning method with DNN}
    \label{fig:cumsum_rate_user_cs3}
\end{figure}

\fref{fig:cumsum_rate_user_cs3} illustrates how the sum rate of each user is distributed with different power allocation algorithms. It is clear that the proposed method gives the highest rate to each user. The figure illustrates that the power allocation with water filling algorithm have high variation of user rate, while the proposed algorithm has a lower variation and higher medium user rate. Thus, we conclude that proposed method gives higher rate for all the users, which results in higher sum rate of the whole network.

\subsubsection*{Note} We can use an auto-encoder to reduce the dimension of the input vector. Thus, reinforcement learning approaches would have a comparatively smaller observation space in comparison when the whole CSI space is utilized as an observation. Also, instead of using DQN we can use policy gradient optimization reinforcement learning approaches that can estimate multiple actions in a single step. This can reduce the number of steps required to estimate a solution (or simply complete an episode). Our method trains it's model using a loss function that does not rely on a labeled dataset. Thus, this can be used as a reward function and use reinforcement learning techniques to train the model optimally. However, these type of reinforcement learning methods are hard to train, therefore we will discuss about the results of these methods in future works.

\section{Conclusion}\label{sec:Conclusion}
This paper presented a comprehensive comparison study on practical design limitation for resource management of learning in non-stationary radio environment.
We studied different problem in single-cell and multi-cell multi-user networks. Through our case studies, we observed specific limitations of learning models such as power violation or general issues such as long-term model ageing that should be taken into consideration in practical design. These limitations make the applicability of learning methods for wireless physical layer problems challenging tasks. We proposed and analyzed solutions to cope with the ageing and power violation problem of learning models. To this end, we highlighted the importance of the unbiased dataset for efficient training of learning model and show that a trade-off between the computational efficiency and prediction accuracy can be balanced with the proper design of the learning architecture. While we proposed the framework for dual and pipeline reinforcement learning approach that efficiently
% \s{not efficient} 
cope with model ageing problem, we also highlight that contemporary learning methods are limited to the applications for non-interactive services.

\begin{IEEEbiography}[{\includegraphics[width=1in,height=1.25in,clip,keepaspectratio]{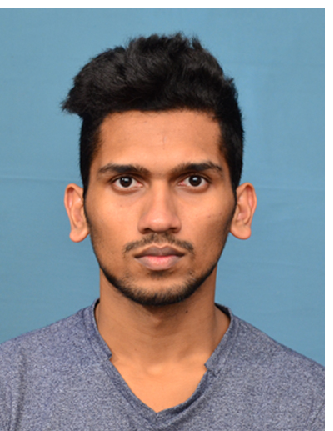}}]{Suren Sritharan} is an undergraduate student in the Computer Engineering program at the University of Peradeniya, Sri Lanka and will be graduating in 2020 with a BS in Computer Engineering. He has a strong interest in the field of Machine Learning, Algorithmic programming and Optimization with applications specifically related to Machine Vision and Wireless Communication.
 \end{IEEEbiography} 

\begin{IEEEbiography}[{\includegraphics[width=1in,height=1.25in,clip,keepaspectratio]{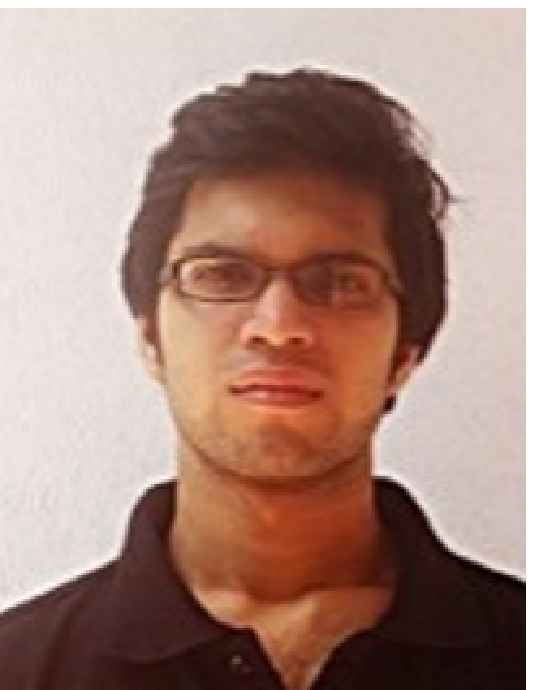}}]{Harshana Weligampola} is pursuing B.Sc. degree on Computer Engineering in University of Peradeninya, Sri Lanka. His research interest include Artificial Intelligence, Neural Networks, Optimization and Image Processing.
 \end{IEEEbiography}

\begin{IEEEbiography}[{\includegraphics[width=1in,height=1.25in,clip,keepaspectratio]{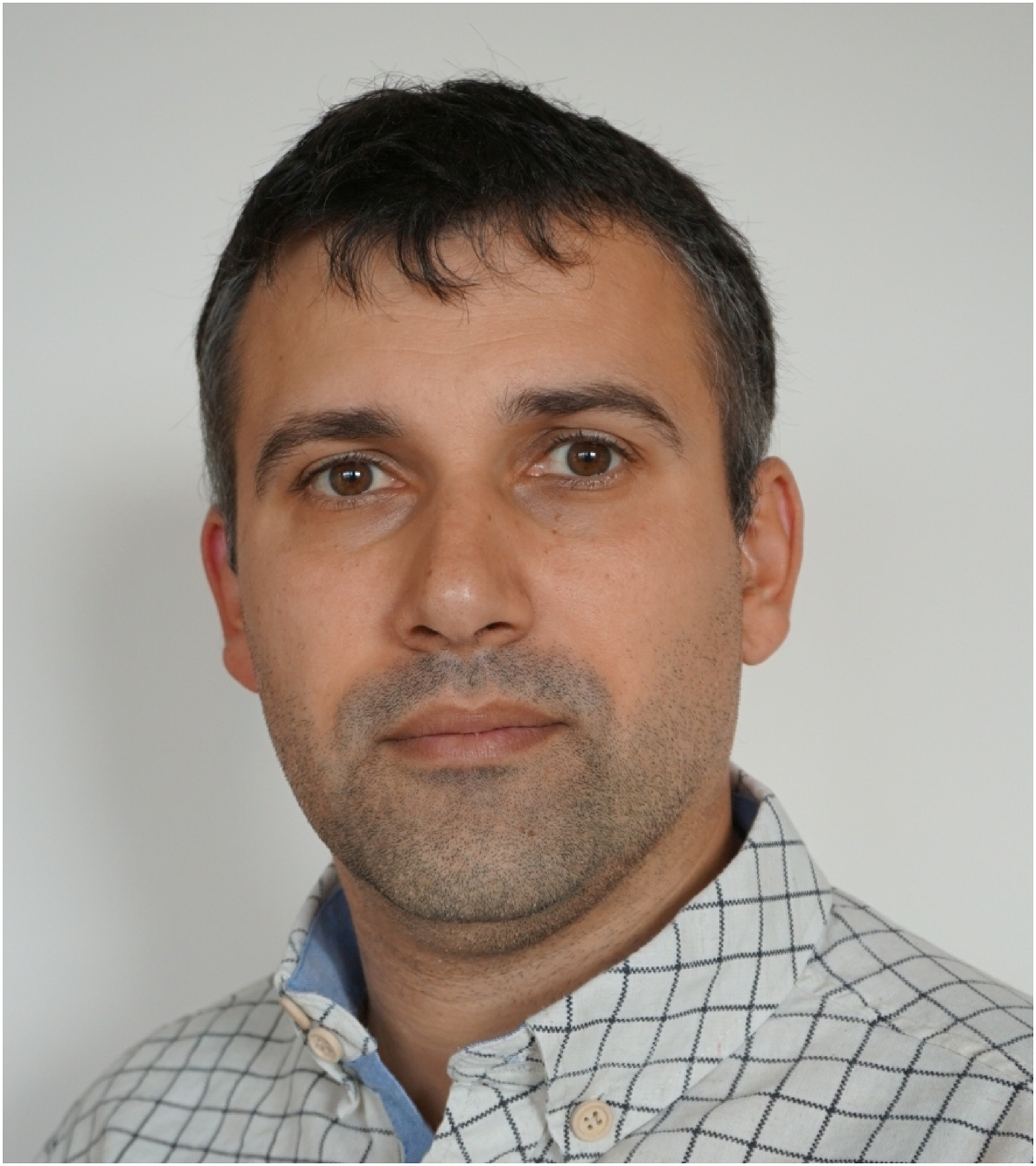}}]{Haris Gacanin} [IEEE SM'12, IEICE SM'12] received his Dipl.-Ing. Degree in Electrical engineering from the University of Sarajevo in 2000. In 2005 and 2008, respectively, he received MSc and Ph.D. from Tohoku University in Japan. He was with Tohoku University from 2008 until 2010 first as Japan Society for Promotion of Science postdoctoral fellow and later, as Assistant Professor. In 2010, he joined Alcatel-Lucent (now Nokia) where he lead research department at Nokia Bell Labs. Currently, he is a full (chair) professor at RWTH Aachen University in Germany. His professional interests are related to broad area of digital signal processing and artificial intelligence with applications in communication systems. He has 200+ scientific publications (journals, conferences and patent applications) and invited/tutorial talks. He is an Associate Editor of IEEE Communications Magazine and previously serviced at IEICE Transactions on Communications and IET Communications. He is IEEE VTS Distinguished Lecturer and he acted as a general chair and technical program committee member of various international conferences. He is a recipient of several Nokia’s awards for innovations, IEICE Communication System Study Group (2015) Award, the 2013 Alcatel-Lucent Award of Excellence, the 2012 KDDI Foundation Research Award, the 2009 KDDI Foundation Research Grant Award, the 2008 Japan Society for Promotion of Science (JSPS) Postdoctoral Fellowships for Foreign Researchers, the 2005 Active Research Award in Radio Communications, 2005 Vehicular Technology Conference (VTC 2005-Fall) Student Paper Award from IEEE VTS Japan Chapter and the 2004 Institute of IEICE Society Young Researcher Award.
 \end{IEEEbiography}

\end{document}